\documentclass[10pt,twocolumn]{article}
\usepackage{graphicx}
\usepackage{amsmath,amssymb,times,cite}

\newcommand{\bm}[1]{\mathbf{#1}} 

\newcommand\T{{\mathpalette\raiseT\intercal}}
\newcommand\raiseT[2]{%
\setbox0\hbox{$#1{#2}$}\raise\dp0\box0}

\usepackage{booktabs}
\usepackage[table]{xcolor}

\definecolor{lightgray}{gray}{0.9}
\definecolor{lightblue}{rgb}{0.93,0.95,1.0}

\title{\Large\textbf{Sharp U-Net: Depthwise Convolutional Network for Biomedical Image Segmentation}}
\author{Hasib Zunair and A. Ben Hamza \\
Concordia Institute for Information Systems Engineering \\
Concordia University, Montreal, QC, Canada
}

\date{}

\begin{document}
\maketitle

\footnotetext[1]{To appear in \textit{Computers in Biology and Medicine}, 2021}

\begin{abstract}
The U-Net architecture, built upon the fully convolutional network, has proven to be effective in biomedical image segmentation. However, U-Net applies skip connections to merge semantically different low- and high-level convolutional features, resulting in not only blurred feature maps, but also over- and under-segmented target regions. To address these limitations, we propose a simple, yet effective end-to-end depthwise encoder-decoder fully convolutional network architecture, called Sharp U-Net, for binary and multi-class biomedical image segmentation. The key rationale of Sharp U-Net is that instead of applying a plain skip connection, a depthwise convolution of the encoder feature map with a sharpening kernel filter is employed prior to merging the encoder and decoder features, thereby producing a sharpened intermediate feature map of the same size as the encoder map. Using this sharpening filter layer, we are able to not only fuse semantically less dissimilar features, but also to smooth out artifacts throughout the network layers during the early stages of training. Our extensive experiments on six datasets show that the proposed Sharp U-Net model consistently outperforms or matches the recent state-of-the-art baselines in both binary and multi-class segmentation tasks, while adding no extra learnable parameters. Furthermore, Sharp U-Net outperforms baselines that have more than three times the number of learnable parameters.
\end{abstract}


\section{Introduction}
Semantic segmentation is a fundamental task in biomedical imaging~\cite{Liangliang:20,LuLi:20,BoWang:20}, with numerous clinical applications including the detection of the novel coronavirus disease 2019 (COVID-19) in computed tomography (CT) images. It refers to the process of classifying each pixel in a biomedical image into one of the pre-defined semantic categories or classes. The goal is to semantically understand the role of each pixel in the image in an effort to distinguish between regions of interests (ROIs) in the image such as tumors or organs~\cite{codella2018skin,yang2018autosegmentation}.

The task of semantic segmentation is tantamount to performing classification at a pixel level, allowing the homogeneous pixels to be clustered together. This motivates the development of automated computer-aided diagnosis systems, which enable physicians to analyze specific image regions, especially in multimodal medical images~\cite{naik2008automated}.

Deep neural networks, and in particular convolutional neural networks (CNNs), have been successfully applied to image segmentation, showing promising results in comparison with shallow methods~\cite{ciresan2012deep,Yuan:17}. Cire\c{s}an \textit{et al.}~\cite{ciresan2012deep} segment biological neuron membranes using a deep neural network as a pixel classifier, where the label of each pixel is predicted from raw pixel values in a sliding window centered around it. Yuan \textit{et al.}~\cite{Yuan:17} present a fully convolutional network for skin lesion segmentation by leveraging a deep convolutional neural network trained end-to-end using Jaccard distance as a loss function. Owing to the recent developments in fully convolutional networks~\cite{long2015fully}, there has been a surge of interest in the adoption of encoder-decoder networks, particularly U-Net, for biomedical image segmentation~\cite{ronneberger2015u,christ2016automatic,cciccek20163d,sirinukunwattana2017gland,zhou2019unet++,kalinin2020medical}. The U-Net model, which is trained in an end-to-end fashion for pixel-wise prediction, has emerged as a very powerful encoder-decoder architecture due largely to its state-of-the-art performance in segmenting biomedical images even when the labelled training data is small. The U-Net architecture is composed of an encoder subnetwork for capturing context by encoding the input image into low-level feature representations at multiple levels and a decoder subnetwork for semantically projecting these feature representations into the pixel space in an effort to enable precise localization via transposed convolutions. While U-Net is built upon the fully convolutional network, it differs from the latter in the sense that it is a symmetric architecture and uses skip connections between the encoder and decoder subnetworks in order to merge low- and high-level features with the goal of preserving more refined image details.

As a core component of encoder-decoder networks, skip connections combine the deep, semantic and course-grained features from the decoder with the shallow, low-level and fine-grained features from the encoder. This feature fusion process has proven to be effective in recovering details of ROIs~\cite{drozdzal2016importance} and in producing accurate segmentation maps, even on a complex image background~\cite{lin2017feature}. While skip connections work relatively well in dense prediction tasks such as image segmentation, the encoder and decoder feature combinations may, however, not match. This mismatch problem is largely attributed to the fact that the encoder features are low-level and fine-grained since they are computed from the early layers of the network, whereas the decoder features are high-level, semantic and course-grained as they are computed from much deeper layers in the network. Consequently, the feature mismatch between the encoder and decoder subnetworks is likely to occur, leading to fusing semantically dissimilar features and hence resulting in blurred feature maps throughout the learning process and also adversely affecting the output segmentation map by under- and/or over-segmenting ROIs.

Prior to fusing with the decoder features, the encoder features undergo a depthwise convolution (i.e. spatial convolution operation performed independently over each channel of the encoder features) by using a sharpening spatial kernel, with the aim to reduce feature mismatch.

To address these limitations, we propose Sharp U-Net, a novel end-to-end depthwise convolutional network architecture for biomedical image segmentation. The key idea behind the proposed framework is to emphasize the fine details of the early level features generated by the encoder via depthwise convolution of the encoder feature map (i.e. spatial convolution operation performed independently over each channel of the encoder features) with sharpening spatial filter prior to performing fusion with the decoder features. Sharpening the features from the encoder subnetwork not only enables better feature fusion, but also help the network progressively learn better representations of the data. Moreover, a sharpening spatial filter helps smooth out artifacts throughout the network during the early stages of training due to untrained parameters. In this paper, we demonstrate that Sharp U-Net yields significantly improved performance over the vanilla U-Net model for both binary and multi-class segmentation of medical images from different modalities, including electron microscopy (EM), endoscopy, dermoscopy, nuclei, and computed tomography (CT). This better performance is achieved without additional learnable parameters in comparison with recent architectures that have more than three times the number of learnable parameters. Finally, the key idea proposed in this work can naturally be used to generalize many other encoder-decoder architectures by incorporating sharpening filters in a similar fashion. To summarize, the main contributions in this paper are as follows:
\begin{itemize}
\item We introduce a novel Sharp U-Net architecture by designing new connections between the encoder and decoder subnetworks using a depthwise convolution of the encoder feature maps with a sharpening spatial filter to address the semantic gap issue between the encoder and decoder features.
\item We show that the Sharp U-Net architecture can be scaled for improved performance, outperforming baselines that have three times the number of learnable parameters.
\item We demonstrate through extensive experiments the ability of the proposed model to learn efficient representations for both binary and multi-class segmentation tasks on a variety of medical images from different modalities.
\end{itemize}

The rest of this paper is organized as follows. In Section 2, we review important relevant work. In Section 3, we outline the motivation behind the proposed framework and present the problem formulation. Then, we introduce an end-to-end depthwise encoder-decoder convolutional network architecture for both binary and multi-class biomedical image segmentation. In Section 4, we present experimental results to demonstrate the competitive performance of our approach on six standard benchmarks. Finally, we conclude in Section 5 and point out future work directions.

\section{Related Work} \label{related}
The task of biomedical image segmentation is to label each pixel of an object of interest in biomedical images, and is often used in clinical applications such as computer-aided diagnosis. Recent variants of the U-Net architecture have focused primarily on improving the performance of U-Net by uniformly scaling the network and/or using pre-trained CNN models on the ImageNet dataset as encoders. Zhou \textit{et al.}~\cite{zhou2019unet++} propose Wide U-Net, which uniformly scales U-Net by increasing the number of filters in both the encoder and decoder subnetworks of U-Net. Also, Zhou \textit{et al.}~\cite{zhou2019unet++} introduce UNet++ , which consists of an ensemble of U-Nets with varying depths and decoders that are densely connected at the same resolution through redesigned skip connections. In spite of improved performance, the UNet++ model is quite complex, requires additional learnable parameters, and some of its components are redundant for specific tasks~\cite{chen2020alpha}. Kalinin \textit{et al.}~\cite{kalinin2020medical} employ ImageNet pre-trained encoders to further improve the performance of U-Net in angiodysplasia lesion segmentation from wireless capsule endoscopy videos and semantic segmentation of robotic instruments in surgical videos. Inspired by Inception modules that are used in CNNs to allow for more efficient computation, Ibtehaz and Rahman~\cite{ibtehaz2020multiresunet} introduce MultiResUNet, an enhanced U-Net architecture, which uses a chain of convolutional layers with residual connections instead of simply concatenating the feature maps from the encoder path to the decoder path. These residual connections not only reduce the semantic gap between the features of the encoder and decoder, but also make the learning easier, while robustly segmenting images from various modalities at different scales.

In light of the availability and quality of medical datasets, there is also a growing interest in developing deep learning frameworks for learning from noisy labels and detecting small anatomical structures with blurry boundaries~\cite{valanarasu2020kiu,mirikharaji2019learning}. Motivated by the performance drop of U-Net in detecting smaller anatomical
structures with blurred noisy boundaries, Valanarasu \textit{et al.}~\cite{valanarasu2020kiu} propose Ki-Net, an over-complete architecture, which projects the data onto high dimensions. When used in conjunction with U-Net, this network yields improved segmentation performance, while having fewer number of parameters. In order to address the problem of detecting small structures with blurry boundaries, Mirikharaji \textit{et al.}~\cite{mirikharaji2019learning} extend the idea of example reweighting in image classification to pixel-level segmentation by training fully convolutional networks from both a large set of weak labels and a small set of expert labels. The idea is to use meta-learning to focus on pixels that have gradients closer to those of expert labels. Ji \textit{et al.}~\cite{ji2020uxnet} employ a neural architecture search based method for volumetric medical image segmentation by searching for scale-wise feature aggregation strategies and blockwise operators in the encoder-decoder network in an effort to generate better feature representations.

While these variants of the U-Net architecture have shown improved results in biomedical image segmentation, the issue of the large semantic gap between the low- and high-level features of the encoder and decoder subnetworks still remains a daunting task. Our work is significantly different from previous work in the sense that the proposed framework mitigates the problem of feature mismatch between the encoder and decoder subnetworks by replacing skip connections with sharpening spatial filters, resulting in much improved segmentation performance. Moreover, our approach can be applied to any encoder-decoder type network.

\section{Proposed Method} \label{Method}
In this section, we start with the motivation behind introducing a depthwise convolution for feature maps sharpening, followed by the problem formulation. Then, we present the main building blocks of our proposed network architecture for binary and multi-class segmentation.

\subsection{Motivation and Problem Statement}
In order to alleviate the fusion of dissimilar features between the encoder and decoder sub-networks of the U-Net model, we extend the U-Net architecture by introducing a depthwise convolution for sharpening the encoder features prior to fusing them with the decoder features. This sharpening operation not only helps balance the semantic gap introduced by the high-level process in the decoder sub-network, but also sharpens the details in the feature maps. In addition, sharpening helps improve the training process in the early stages by reducing low-frequency noise propagated by the untrained parameters in the feature space. It is important to point out that our Sharp U-Net architecture does not introduce any additional learnable parameters.

\medskip\noindent\textbf{Problem Formulation.}\quad Semantic segmentation refers to the process of classifying each pixel in an image into its semantic class and hence can be regarded as a classification problem at the pixel level. A lung segmentation task, for example, can be thought of as a binary segmentation problem with two semantic classes: lung and background. However, unlike image classification whose goal is to assign an input image to one label from a fixed set of categories or classes, the output in semantic segmentation is an image, typically of the same size as the input image, such that each pixel is classified to a particular class.

For a multi-class segmentation problem consisting of $C$ classes, we denote by $\mathcal{X}=\{(\bm{x}_i,y_i):\,i=1,\dots,N\}$ a set of $N$ samples, where $\bm{x}_i$ is the $i$th training sample and $y_{i}\in\{1,\dots,C\}$ is the corresponding true label. The true label of the $i$th sample can be represented as a one-hot encoding vector $\bm{y}_{i}=(y_{i1},\dots,y_{iC})^\T$, such that $y_{ic}=1$ if $i=c$ and 0 otherwise for each class $c$.

\subsection{Proposed Neural Architecture}
\begin{figure*}[!htb]
\centering
\includegraphics[scale=.55]{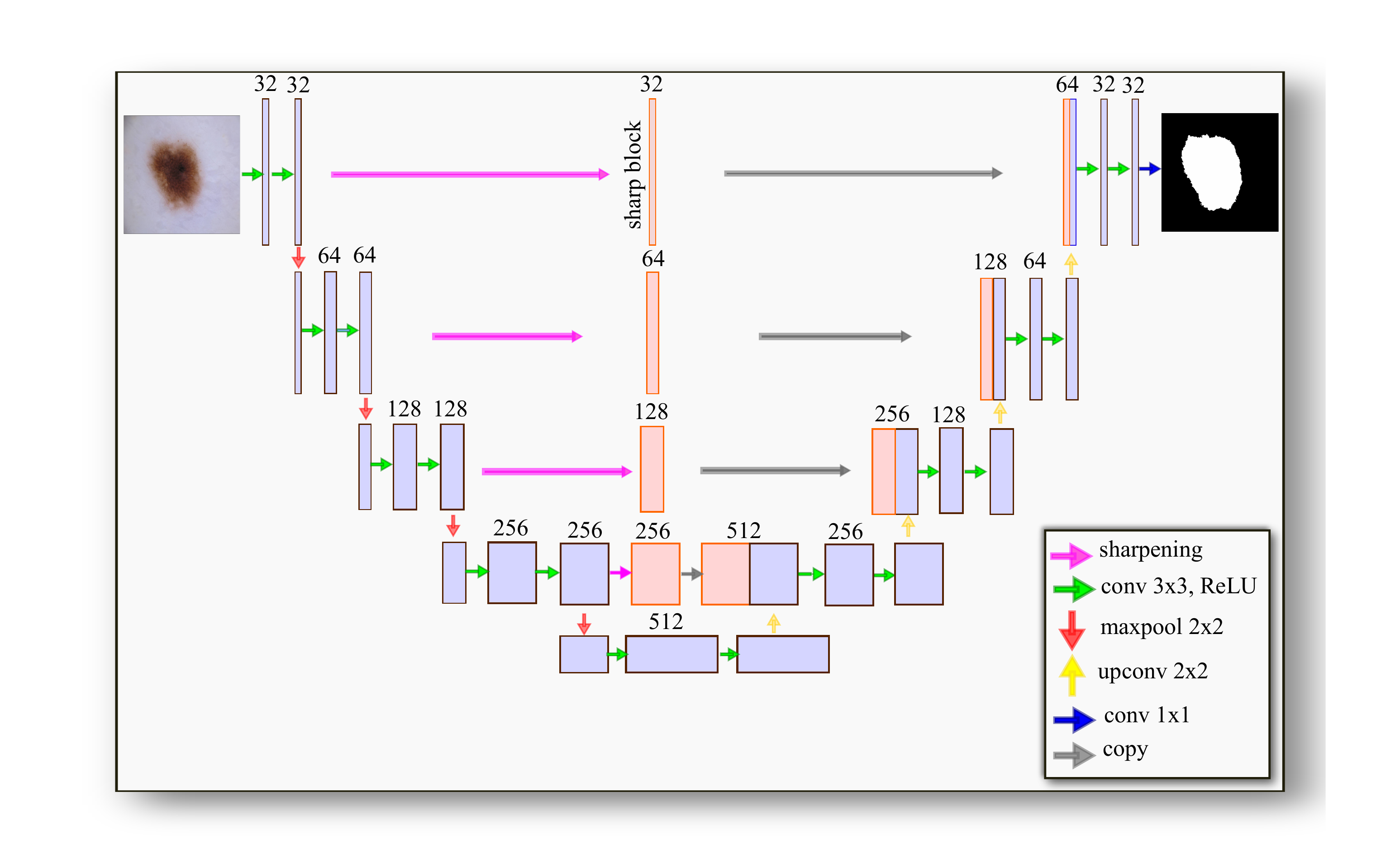}
\caption{Schematic layout of the proposed Sharp U-Net architecture. Prior to fusing with the decoder features, the encoder features undergo a depthwise convolution (i.e. spatial convolution operation performed independently over each channel of the encoder features) by using a sharpening spatial kernel, with the aim to reduce feature mismatch. These additional operations do not increase the number of learnable parameters, and hence no additional computational cost is incurred.}
\label{Fig:architecture}
\end{figure*}

Skip connections in encoder-decoder networks, particularly in U-Net, play a crucial role in recovering fine-grained details in the prediction. However, these skip connections tend to fuse low- and high-level convolutional features that are semantically different, and hence resulting in blurred feature maps. In order to address these issues, we introduce a Sharp U-Net architecture for both binary and multi-class segmentation, as shown in Figure~\ref{Fig:architecture}. The key rationale of our Sharp U-Net
framework is to mitigate the semantic gap between the encoder and decoder features, and in the meanwhile, to smooth out artifacts throughout the network layers during the early stages of training. Similar to U-Net, the proposed Sharp U-Net architecture consists of a contracting or downsampling path (encoder) for capturing context using convolutions and an expanding or upsampling path (decoder) for enabling precise localization using transposed convolutions (also known as up-convolutions).

The encoder, which is composed of five blocks, down-samples the input feature maps to extract low-level features. Each block of the encoder consists of two $3\times 3$ convolutional layers with rectified linear unit (ReLU) activations, followed by a $2\times 2$ max-pooling layer, except the fifth block, which does not include a pooling layer. We use 32, 64, 128, 256 and 512 filters for the convolutional layers in the encoder blocks, i.e. the number of features maps is doubled after each block. The convolutional layer applies a filter to an input to create a feature map that summarizes the presence of extracted features from the input, whereas the pooling layer downsamples or reduces the size of each feature map by a factor of 2.

On the other hand, the decoder, which also consists of five blocks, may be regarded as an operator that performs the reverse of the downsampling path. Every block in the expanding path comprises a $2\times 2$ up-convolution (i.e. upsampling the features), followed by two $3\times 3$ convolutions with ReLU activations, except the fifth block, which has an additional $1\times 1$ convolution. We use 256, 128, 64 and 32 filters for the convolutional layers in the decoder blocks, i.e. the number of features maps is halved after each block. The output of the Sharp U-Net model is a pixel-by-pixel mask that shows the class of each pixel.

In order to localize the upsampled features, we design a new connection mechanism, called sharp block, between the encoder and decoder subnetworks in an effort to fuse the low- and high-level features from the encoder and decoder, respectively, while mitigating the semantic gap problem. To this end, instead of using plain skip connections between the encoder and decoder, the encoder features undergo a spatial convolution operation conducted independently on each channel of the encoder features using a sharpening spatial kernel prior to performing fusion with the decoder features. This sharpening filter layer not only fuses semantically less dissimilar features, but also helps reduce the high-frequency noise components that are propagated throughout the network layers during the early stages of training as the channels are Gaussian smoothed before applying the Laplacian filter. Moreover, scaling strategies can be leveraged to further improve the performance of the proposed Sharp U-Net framework by uniformly increasing the learnable parameters and/or using pre-trained CNN models as encoders.

\subsection{Sharpening Spatial Kernel} \label{sharp}
Spatial filtering is a low-level neighborhood-based image processing technique, which performs operations on the neighborhood of every pixel in an image for the sake of image enhancement such as sharpening. The objective of sharpening spatial filtering is to preserve or emphasize high-frequency components representing fine-grained image details by highlighting transitions in intensity within the image. The image sharpening process, also referred to as high-pass filtering, is usually performed via image convolution with kernels or masks. Convolution kernels, also known as filters, are discrete approximations of the image Laplacian, which is a second-order derivative operator that is capable of responding to intensity transitions in any direction. A commonly-used Laplacian high-pass filtering kernel for image sharpening, which takes into account all eight neighbors of the reference pixel in the input image, is given by
\begin{equation}
\bm{K}=\begin{bmatrix}
    -1 & -1 & -1\\
    -1 &  8 & -1\\
    -1 & -1 & -1
\end{bmatrix}.
\end{equation}
Note that convolving an image with the Laplacian filter kernel $\bm{K}$ increases the brightness of the center pixel relative to neighboring pixels, as the $3\times 3$ kernel matrix is comprised of a positive value in the center and negative off-center values. Also, the sum of all elements of the high-pass filter kernel is always zero.

In order to obtain a sharpened image, the input image is added to its convolution with the kernel. More precisely, let $\bm{I}$ be the input image, then the resulting sharpened image $\bm{S}$ is given by
\begin{equation}
\bm{S} = \bm{I} + \bm{K}\ast\bm{I},
\end{equation}
where $\ast$ denotes convolution, a neighborhood-based operator that processes an image by adding each value of a pixel to its local neighbors, weighted by the kernel.

\subsection{Sharp Blocks}
Since the encoder features are multi-dimensional, generally of size $W\times H\times M$ with $W$, $H$ and $M$ denoting the width, height and number of encoder feature maps respectively, we perform a depthwise convolution on each feature map using a sharpening spatial kernel given by the Laplacian filter kernel $\bm{K}$. Similar to a kernel in a convolutional layer, the depthwise convolution layer is parametrized by the sharpening spatial kernel $\bm{K}$. This can also be thought of as initializing the weights of the depthwise convolutional layer with the weights given by $\bm{K}$. It is worth pointing out that the Laplacian spatial filter kernel has no learnable parameters. Hence, no new parameters are updated during model optimization; thereby no additional computation cost is incurred.

\begin{figure}[!htb]
\centering
\includegraphics[scale=.45]{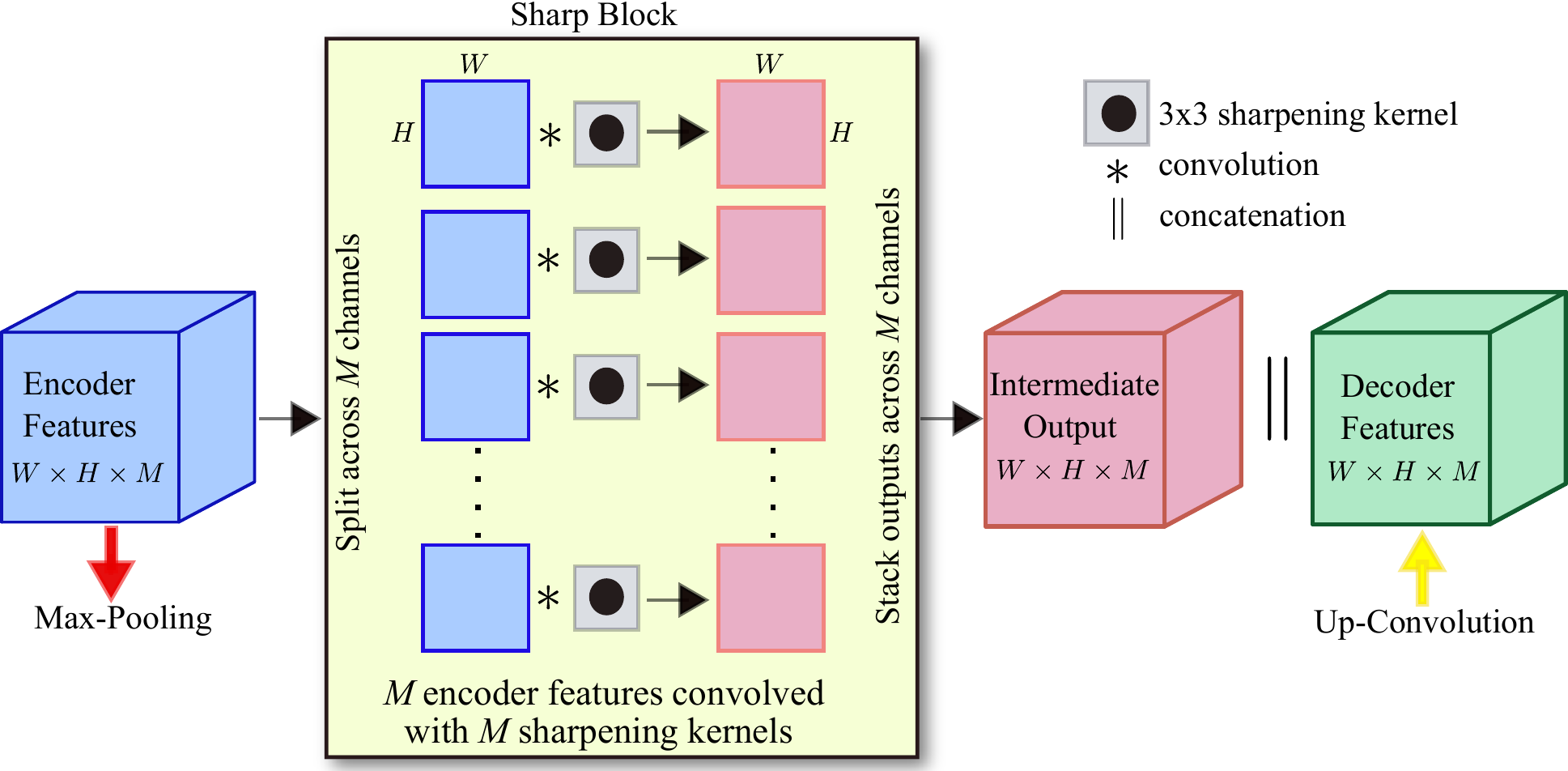}
\caption{Illustration of the proposed Sharp Block. Given a multi-channel encoder feature map of size $W\times H\times M$ as input, the sharpening kernel layer performs a depthwise convolution, resulting in a sharpened intermediate feature map of the same size as the input.}
\label{Fig:sharpath}
\end{figure}

Since we are interested in transforming the input encoder features, no additional bias terms are used for this initialization. Instead of using a single filter of a particular size (i.e. $3\times 3\times 3$) in convolutions, we use $M$ filters, which act on each input channel separately. Each channel of the input is convolved with the kernel $\bm{K}$ separately with a stride of 1. Each of these convolutions yields a feature map of size $W\times H\times 1$. During the feature fusion of the encoder and decoder subnetworks, we also perform padding to keep the output dimension the same as that of the input so that it matches the size of the decoder features in all stages of the connection. We then stack these maps together to obtain the final output of size $W\times H\times M$ from the depthwise convolution layer. We refer to this proposed feature connection as a sharp block. A visualization of the flow of operations in the sharp block is depicted in Figure~\ref{Fig:sharpath}.

\subsection{Scaling Sharp U-Net}
The proposed Sharp U-Net architecture can be scaled using recent scaling strategies such as uniformly increasing the number of learnable parameters~\cite{zhou2019unet++}, and/or using deep convolutional neural networks in the encoder subnetwork~\cite{kalinin2020medical,he2016deep}. Uniformly increasing the number of learnable parameters is achieved by increasing the number of convolutional kernels while maintaining the kernel size. We increase the number of kernels of both the encoder and decoder subnetworks from 32, 64, 128, 256 and 512 to 35, 70, 140, 280 and 560, respectively. We refer to the resulting network architecture as Wide Sharp U-Net. We also scale Sharp U-Net by replacing the encoder with a pretrained convolutional neural network (e.g., VGG~\cite{kalinin2020medical} or ResNet~\cite{he2016deep}) on the 1000-class Imagenet dataset, as the encoder plays the role of a feature extractor. In the different configurations of the scaled Sharp U-Net, an increase in the number of learnable parameters comes from the scaling strategies.

\subsection{Model Training} \label{optim}
Similar to the U-Net architecture, the proposed Sharp U-Net for multi-class segmentation is trained in an end-to-end fashion. The parameters (i.e. weights and biases for different layers) of the Sharp U-Net model are learned by minimizing the following categorical cross-entropy loss for all $N$ samples
\begin{equation}
\mathcal{L}=-\sum_{i=1}^{N}\sum_{c=1}^{C}y_{ic}\log\hat{y}_{ic},
\end{equation}
where $C$ is the total number of classes, $y_{ic}$ is the indicator that the $i$th sample belongs to the $c$th class, and $\hat{y}_{ic}$ is the predicted probability that the model associates the $i$th input with class $c$. During training, the network parameters are updated using the Adam optimizer~\cite{kingma2014adam}. Training is continued on all network layers until the validation loss stops improving, and then the best weights are retained using early an stopping mechanism. It is worth mentioning that other loss functions such as the focal Tversky loss~\cite{abraham2019novel} and distance-based losses~\cite{karimi2019reducing, caliva2019distance} can also be used for training.

\section{Experiments} \label{exps}
In this section, we conduct extensive experiments to assess the performance of the proposed image segmentation framework in comparison with strong baseline methods on several benchmark datasets. The source code to reproduce the experimental results will be made publicly available on GitHub\footnote{https://github.com/hasibzunair/sharp-unets}.

\subsection{Datasets} \label{datasets}
We demonstrate and analyze the performance of the proposed image segmentation model on four datasets: Lung Segmentation\footnote{https://www.kaggle.com/kmader/finding-lungs-in-ct-data}, Data Science Bowl 2018 segmentation challence\footnote{https://www.kaggle.com/c/data-science-bowl-2018}, ISIC-2018, COVID-19 CT Segmentation\footnote{https://medicalsegmentation.com/covid19}, ISBI-2012~\cite{cardona2010integrated}, and CVC-ClinicDB~\cite{bernal2015wm}. We use the COVID-19 CT Segmentation dataset for multi-class image segmentation and the remaining datasets for binary image segmentation. The summary descriptions of these benchmark datasets are as follows:
\begin{itemize}
	
\item \textbf{Lung Segmentation} is a collection of 2D and 3D CT images with manually segmented lungs. The image size is $512\times 512$, with variable depth. In our experiments, we use the slices of the CT images and their corresponding masks. For data preprocessing, we follow the same procedure as~\cite{azad2019bi}.

\item \textbf{Data Science Bowl 2018} is a dataset comprised of 670 segmented nuclei images from different modalities, brightfield and fluorescence in particular. This dataset also consists of instance-level annotations, where each cell is marked with a unique color or label. These 670 images are of various resolution, and we resize them to $256\times 256$ while maintaining the aspect ratio.

\item \textbf{ISIC-2018} is a dataset composed of skin lesions with a total of 2594 images with both segmentation masks expert labels from the HAM10000 dataset~\cite{tschandl2018ham10000} and ISIC-2017 dataset~\cite{codella2018skin}. The dataset consists of images of various resolutions, which we resize them to $256\times 192$ while retaining aspect ratio.

\item \textbf{COVID-19 CT Segmentation} is a multi-class segmentation dataset, which consists of 100 axial CT images from about 40 patients with COVID-19. These images were segmented by a radiologist using three labels: ground-glass, consolidation and and pleural effusion. All images are of size $512\times 512$.

\item \textbf{ISBI-2012} is an electron microscopy image modality dataset, which comprises 30 images from a serial section Transmission electron Microscopy (ssTEM) of Drosophila first instar larva ventral nerve cord XX. The images consist of alignment errors and also have noisy examples. We resize these images to $256\times 256$.

\item \textbf{CVC-ClinicDB} is a colonoscopy image dataset, which consists of endoscopy images. These images were extracted from video sequences of colonoscopy. Since only images of polyp are considered for the task of segmentation, this results in a total of 612 images. We resize these images to $256\times 192$ while preserving the aspect ratio while maintaining the aspect ratio.
\end{itemize}

\subsection{Baselines} \label{baselines}
We evaluate the performance of the proposed method against various baselines, including U-Net~\cite{ronneberger2015u}, Wide U-Net~\cite{zhou2019unet++}, TernausNet-16~\cite{kalinin2020medical}, and U-Net + ResNet-50~\cite{he2016deep}. We further improve these networks using our proposed architecture to demonstrate the extensibility of our approach, while avoiding additional learnable parameters. For baselines, we mainly consider methods that are closely related to U-Net and/or the ones that are state-of-the-art biomedical image segmentation frameworks. A brief description of these strong baselines can be summarized as follows:

\begin{itemize}
\item \textbf{U-Net} is a fully convolutional network for binary and multi-class biomedical image segmentation. It consists of a contracting path to capture context and a symmetric expanding path that enables precise localization. The five convolutional layers in the contracting path consist of 32, 64, 128, 256 and 512 filters, while the convolutional layers in the expanding path are comprised of 256, 128, 64 and 32 filters. For an input image of size $192\times 256$, the U-Net model has approximately 7.8 million learnable parameters.

\item \textbf{Wide U-Net} is a scaled version of U-Net, and is obtained by increasing the number of filters in both contracting and expanding paths (i.e. encoder and decoder). The convolutional layers in the contracting path have 35, 70, 140, 280 and 560 filters. The same numbers of filters are used in the expanding path. The network has roughly 9.1 million learnable parameters.

\item \textbf{TernausNet-16} is a variant of U-Net, in which the contracting path is replaced with a VGG-16 pretrained model on ImageNet. The VGG-16 model consists of 16 layers with learnable weights: 13 convolutional layers, and 3 fully connected layers. Each of the first and second convolutional blocks is comprised of two convolutional layers with 64 and 128 filters, respectively.  Similarly, each of the third, fourth and fifth convolutional blocks consists of three convolutional layers with 256, 512, and 512 filters, respectively. The network has about 23.7 million learnable parameters.

\item \textbf{U-Net + ResNet-50} is a segmentation network, where the encoder part of the U-Net is replaced with the ResNet-50 pretrained model on ImageNet. The network has approximately 32.5 million learnable parameters.
\end{itemize}

It is important to note that in all experiments, no data augmentation or post-processing such as conditional random fields (CRF) or median filtering are employed, as our aim is to introduce a new architecture and demonstrate the performance improvements attributed to sharp blocks.

\subsection{Implementation Details} \label{implementation}
All experiments are run on a Linux Workstation featuring an AMD Ryzen Threadripper 2950X processor with 16 cores and 64 processing threads, 4.4 GHz Max Boost, 64GB RAM, and an NVIDIA GeForce RTX 2080 Ti with 11 GB Memory. We use $k$-fold cross-validation with $k=5$ to evaluate the segmentation results by different methods using two evaluation metrics. The cross-validation experiments are conducted 5 times with different random splits of the data. The average and standard deviation scores are reported. We use the Adam optimizer with learning rate 0.001.

\subsection{Evaluation Metrics}
In order to evaluate the performance of our proposed framework against the baseline methods, we use the Jaccard index and Dice coefficient as evaluation metrics. The Jaccard index, also known as Intersection-Over-Union (IoU), is one of the most commonly used metrics in semantic segmentation. Given two sets $G$ and $P$, denoting the ground truth binary and predicted labels, respectively, the Jaccard similarity index is defined as
\begin{equation}
\mathcal{J}(G,P) = \frac{|G\cap P|}{|G\cup P|},
\end{equation}
where $|\cdot|$ denotes the cardinality of a set. Thus, the Jaccard index is the area of overlap between the predicted segmentation and the ground truth divided by the area of union between the predicted segmentation and the ground truth. For binary or multi-class segmentation, the mean IoU is computed by taking the IoUs of all classes and averaging them. The Jaccard index ranges from 0 to 1, with 1 indicating perfect match between the true and predicted labels, while 0 indicates a complete mismatch between them.

The Dice similarity coefficient is defined as the area of overlap between the predicted segmentation and the ground truth divided by the average of the areas of the predicted segmentation and the ground truth:
\begin{equation}
\mathcal{S}(G,P) = \frac{2|G\cap P|}{|G|+|P|}.
\end{equation}
The Dice similarity coefficient also has values in the range [0, 1]. A similarity of 1 means that the segmentations are a perfect match.
\subsection{Image Segmentation Performance}
We start by showing model validation history comparison between U-Net and the proposed Sharp U-Net model on all datasets. Then, we provide experimental results for both binary and multi-class segmentation on biomedical images from multiple modalities. We also show qualitative results of the predictions on some hard examples.

\begin{figure}[!htb]
\setlength{\tabcolsep}{.2em}
\centering
\begin{tabular}{cc}
\includegraphics[scale=.31]{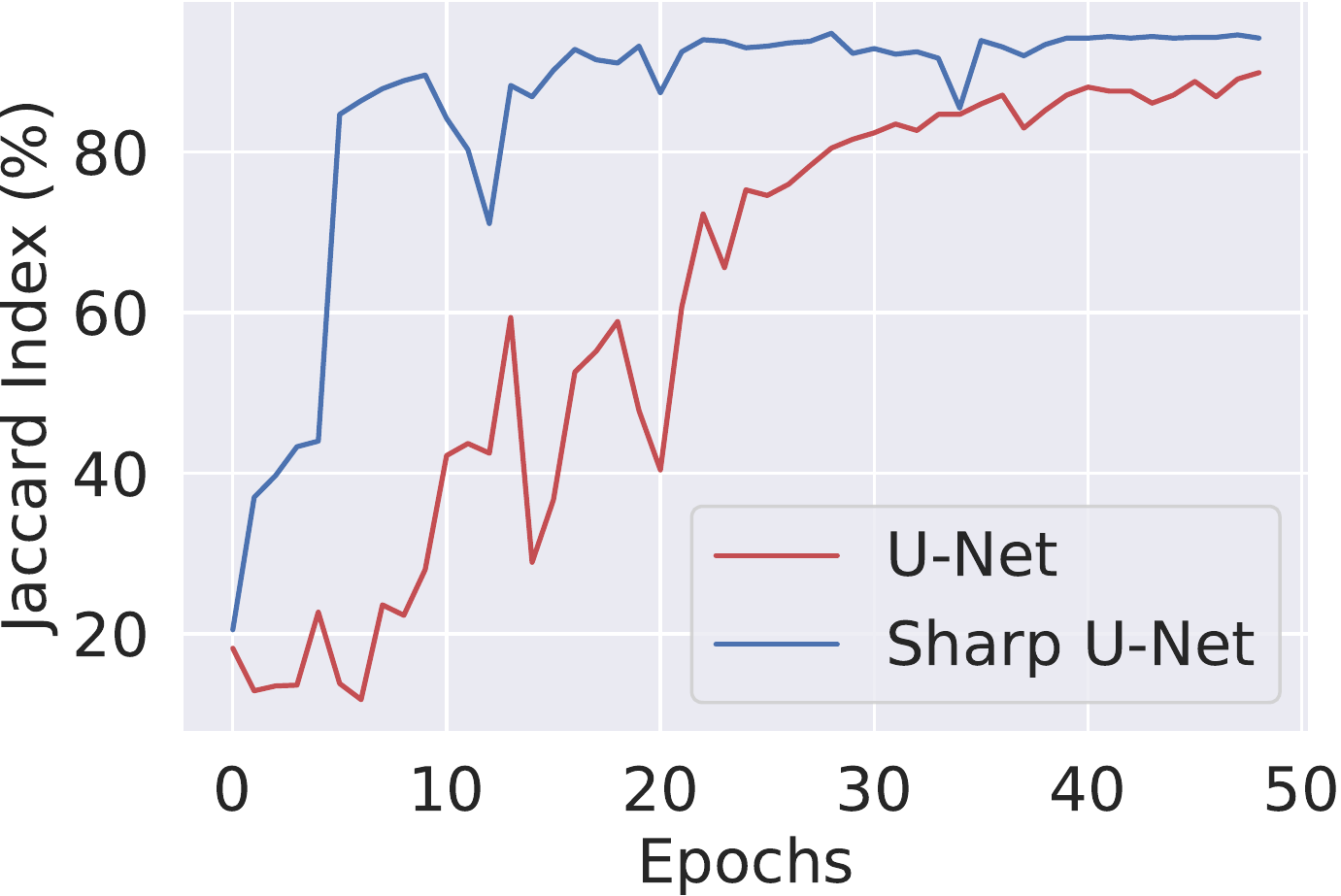} & \includegraphics[scale=.31]{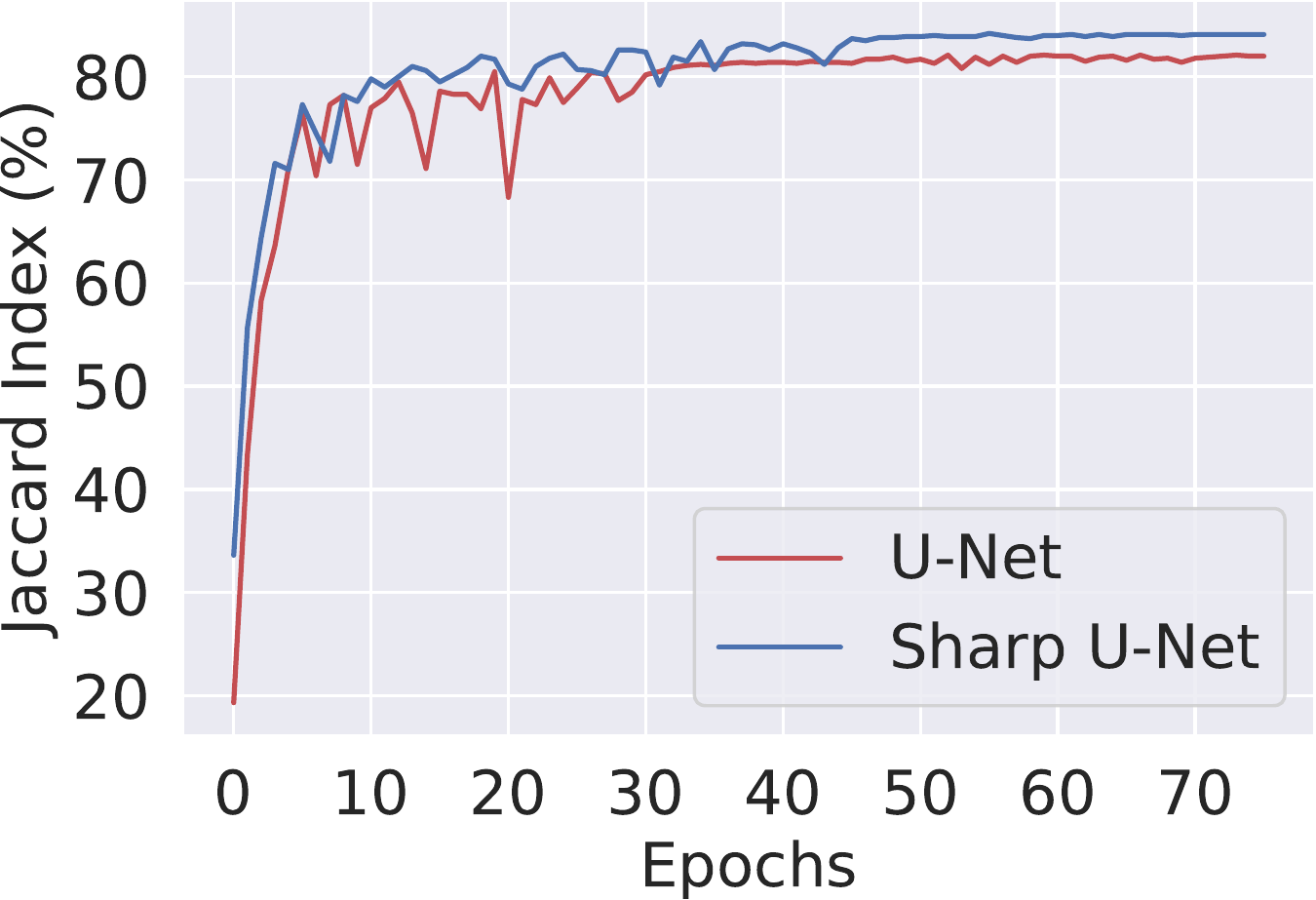} \\
(a)& (b)\\[1ex]
\includegraphics[scale=.31]{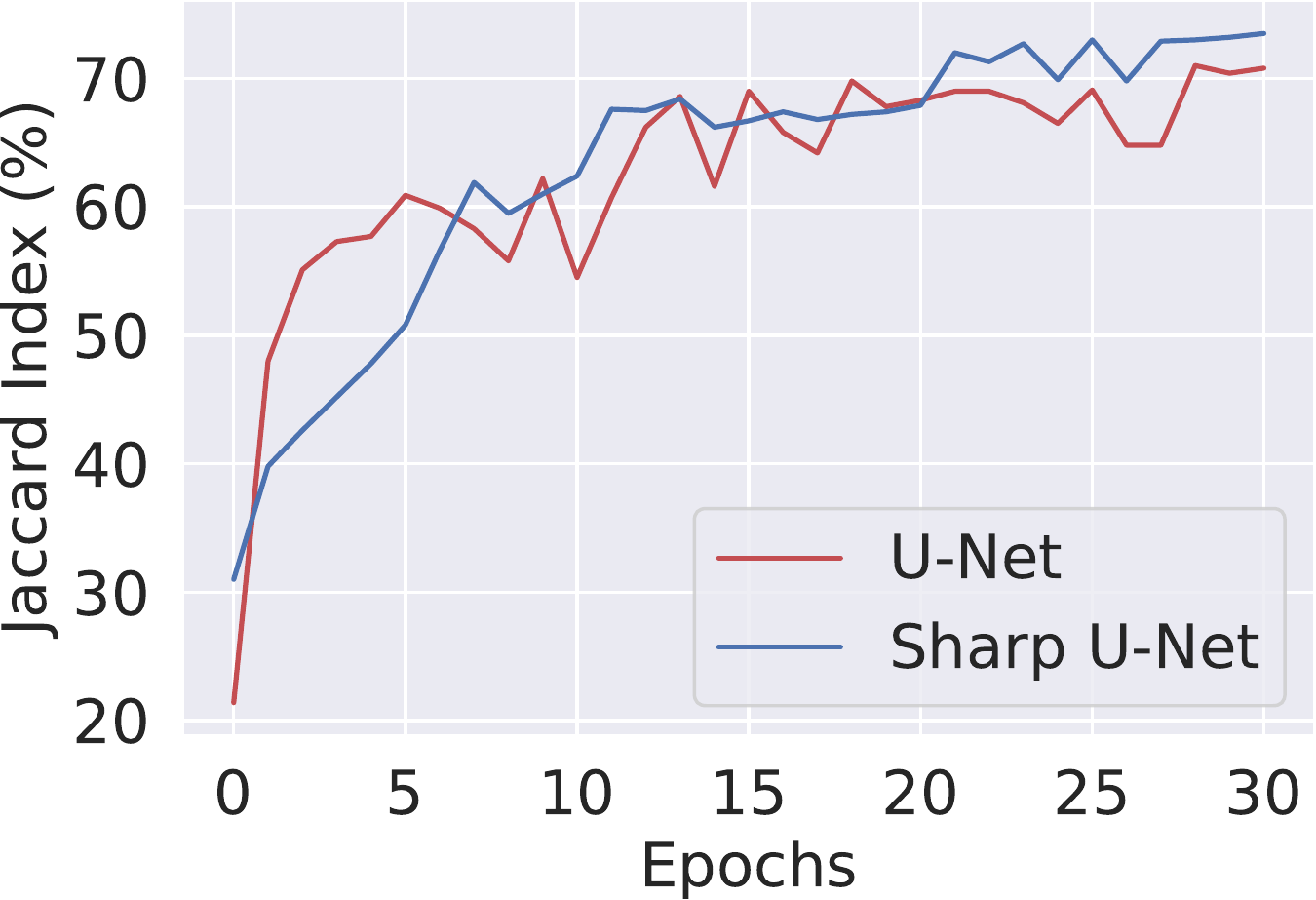} & \includegraphics[scale=.31]{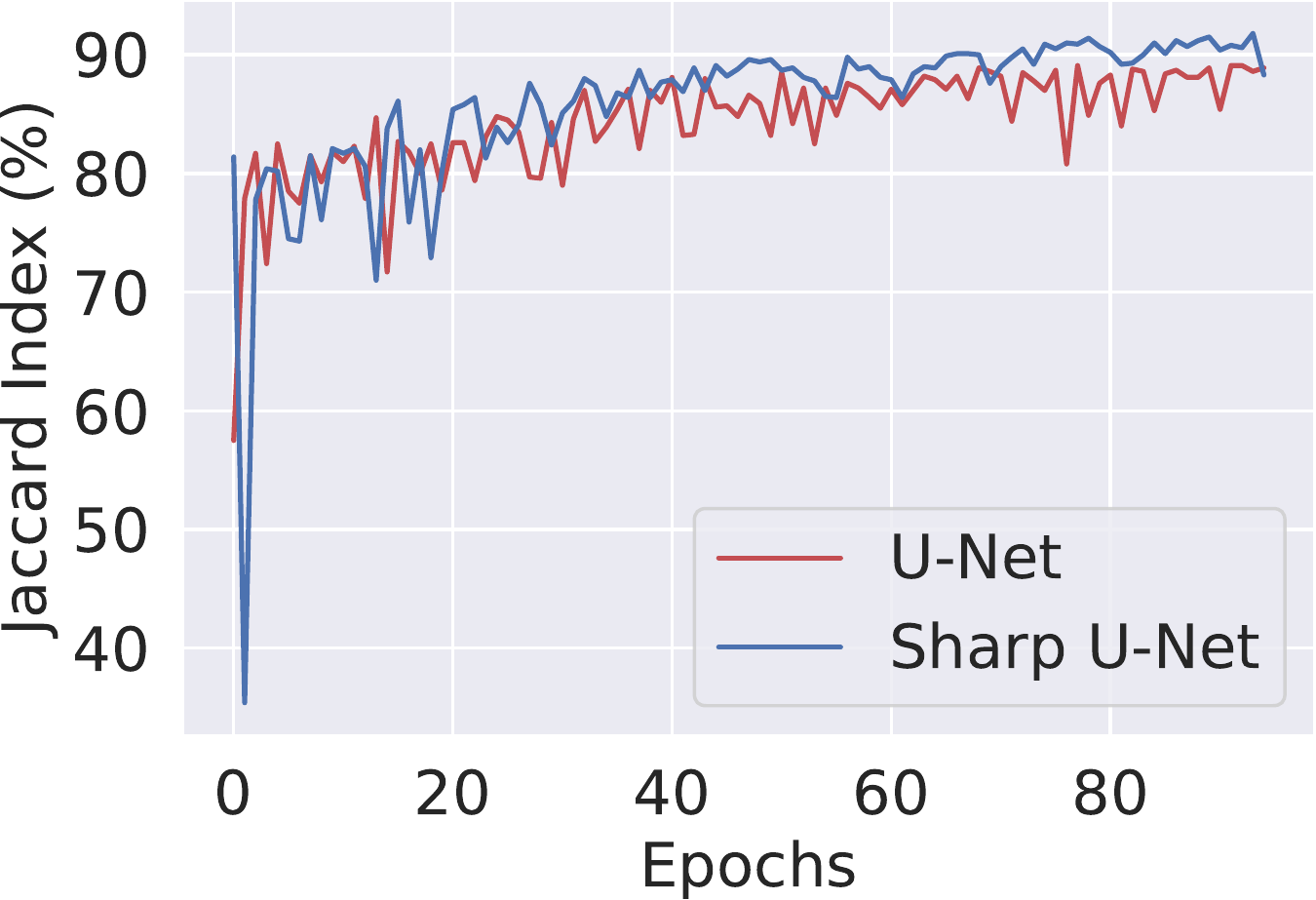}\\
(c) & (d) \\ [1ex]
\includegraphics[scale=.31]{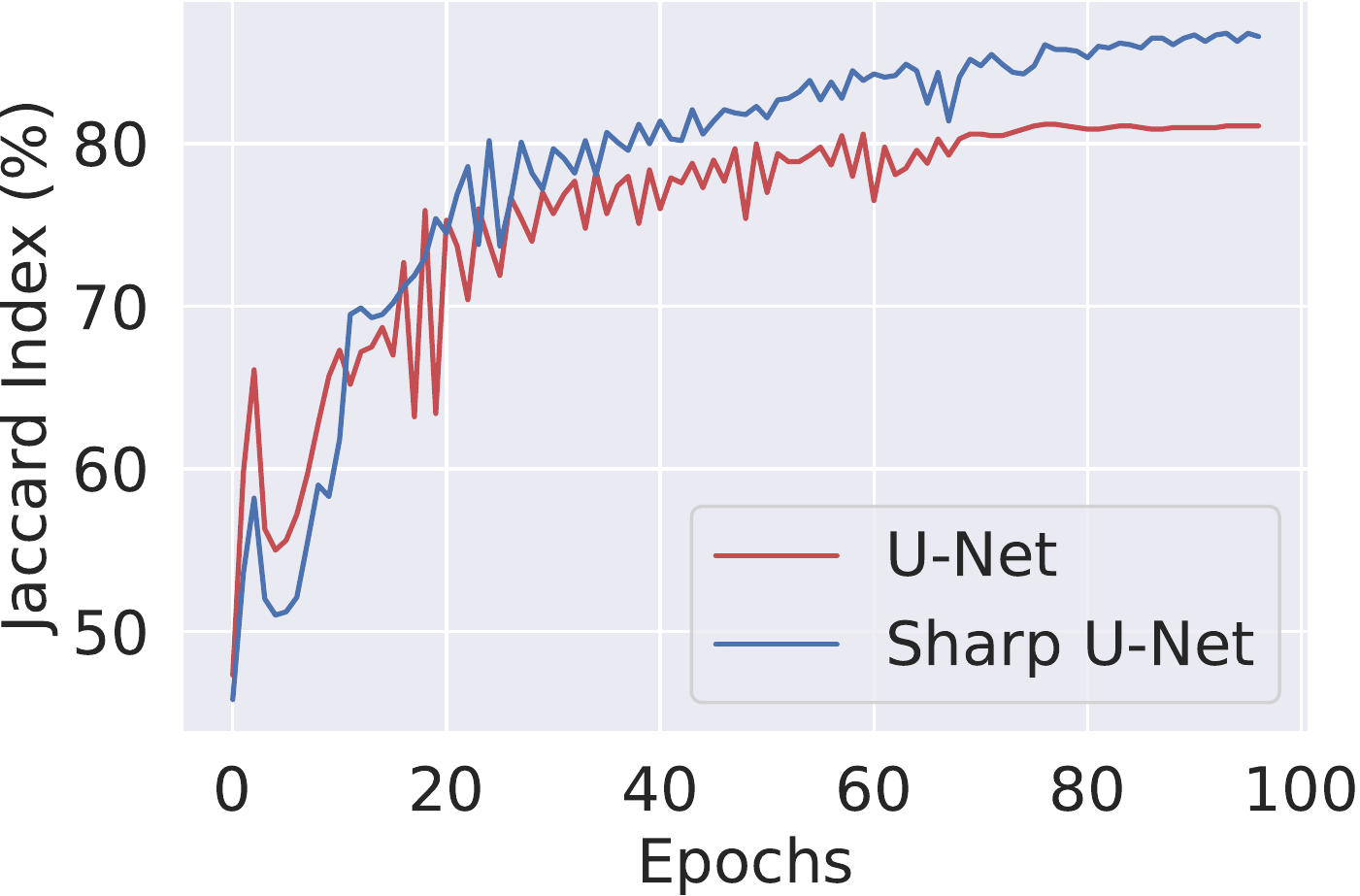} & \includegraphics[scale=.31]{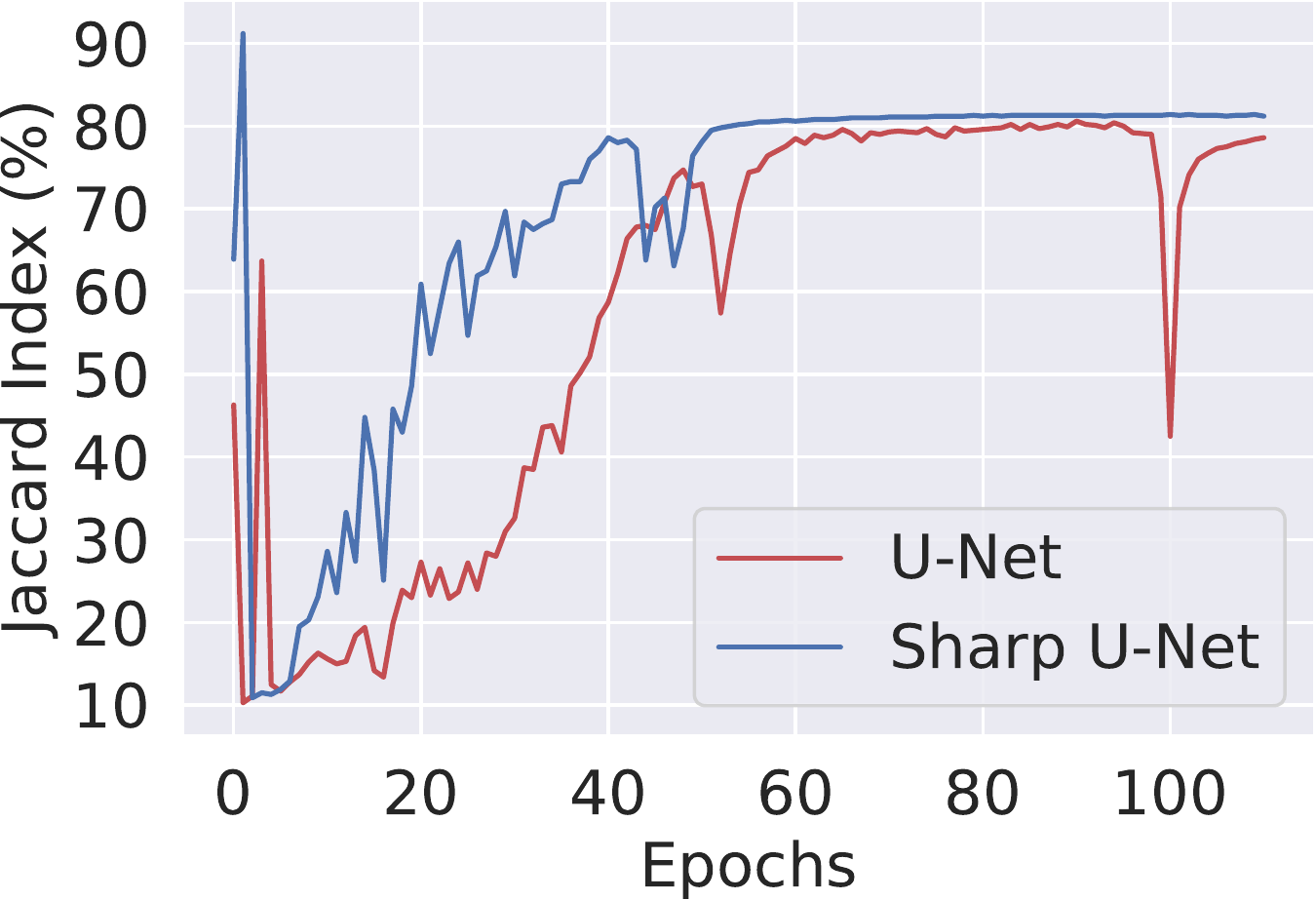}\\
(e) & (f)
\end{tabular}
\caption{Progress of the best validation performance with the number of epochs from 5-fold cross-validation tests: (a) Lung Segmentation, (b) Data Science Bowl 2018, (c) ISIC-2018, (d) COVID-19 CT Segmentation, (e) ISBI-2012, and (f) CVC-ClinicDB. We record the value of the Jaccard index on the validation set after each epoch.}
\label{fig:graphs}
\end{figure}

\medskip\noindent\textbf{Model validation performance.}\quad The performance comparison between U-Net and Sharp U-Net over validation epochs on the validation set of each dataset is illustrated in Figure~\ref{fig:graphs}, which shows that the proposed Sharp U-Net model yields higher Jaccard index scores, indicating better performance. As can be seen, Sharp U-Net achieves better performance using 5-fold cross-validation on all cases for the same number of epochs. This better performance is largely attributed to the prevention of noise that is propagated during the early stages of training due to untrained parameters. In Figure~\ref{fig:graphs}(a), we can see that Sharp U-Net achieves better results much faster than U-Net. In Figure~\ref{fig:graphs}(f), both U-Net and Sharp U-Net display major fluctuations during training, but Sharp U-Net converges much faster, whereas U-Net lags behind in the early stages of the training process. These results demonstrate that Sharp U-Net yields superior results using the same number of epochs and even faster in some cases compared to the U-Net model.

\medskip\noindent\textbf{Sharp U-Net outperforms U-Net.}\quad We evaluate the performance of Sharp U-Net in comparison with the U-Net baseline across various datasets consisting of multiple modalities for both binary and multi-class segmentation tasks. For each task, we perform 5-fold cross-validation tests. The best results in each run are recorded and then averaged across all runs to get the scores of the Jaccard Index and Dice for each model. We report the results in Table~\ref{Tab:results1}, which shows that the proposed Sharp U-Net model performs better than the U-Net architecture in segmenting medical images from different modalities. Significant performance improvements are observed in the case of endoscopy and EM images. A relative improvement of 12.6\% and 3.63\% on the Jaccard Index is observed on the CVC-ClinicDB and ISBI-2012 datasets. Note that this performance improvement is achieved with no additional learnable parameters. For dermoscopy images, Sharp U-Net achieves a relative improvement of 1.79\% on the ISIC-2018 dataset. For CT and Nuclei images, Sharp U-Net matches the performance of U-Net, albeit Sharp U-Net performs slightly better with relative improvements of 0.48\% and 0.63\% on the Lung Segmentation and Data Science Bowl 2018 dataset, respectively. A similar pattern can be observed with the Dice score evaluations.

\begin{figure*}[!htb]
\centering
\includegraphics[width=5.5in, height=1.7in]{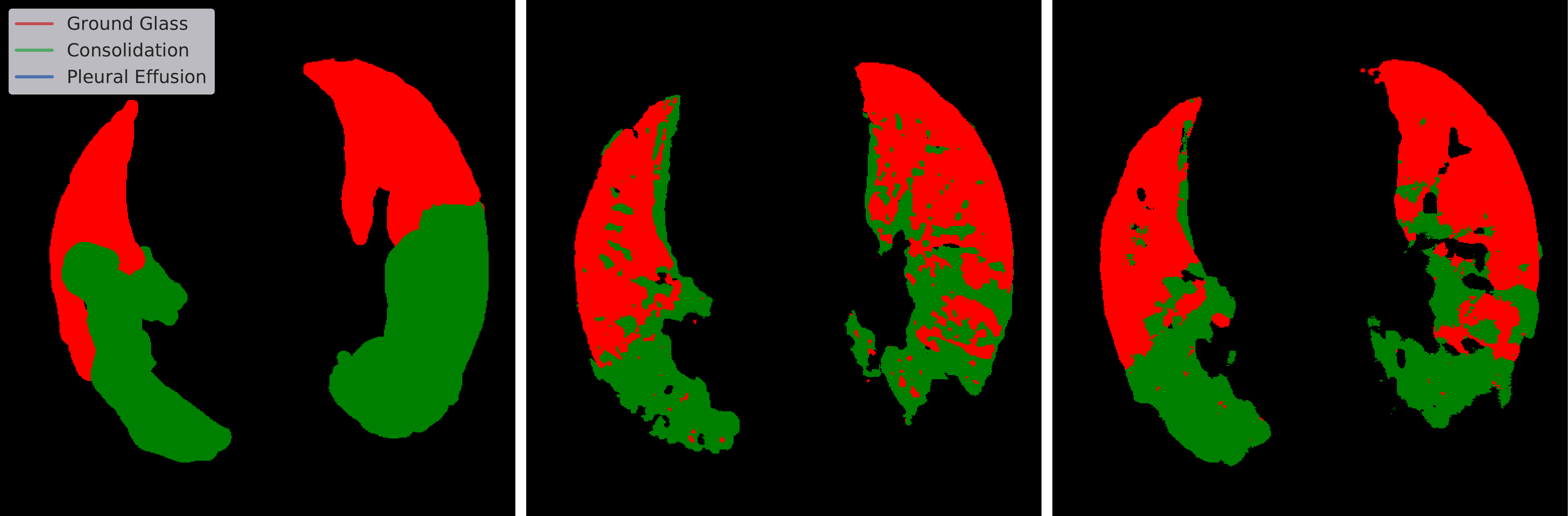}\\[.6ex]
\includegraphics[width=5.5in, height=1.7in]{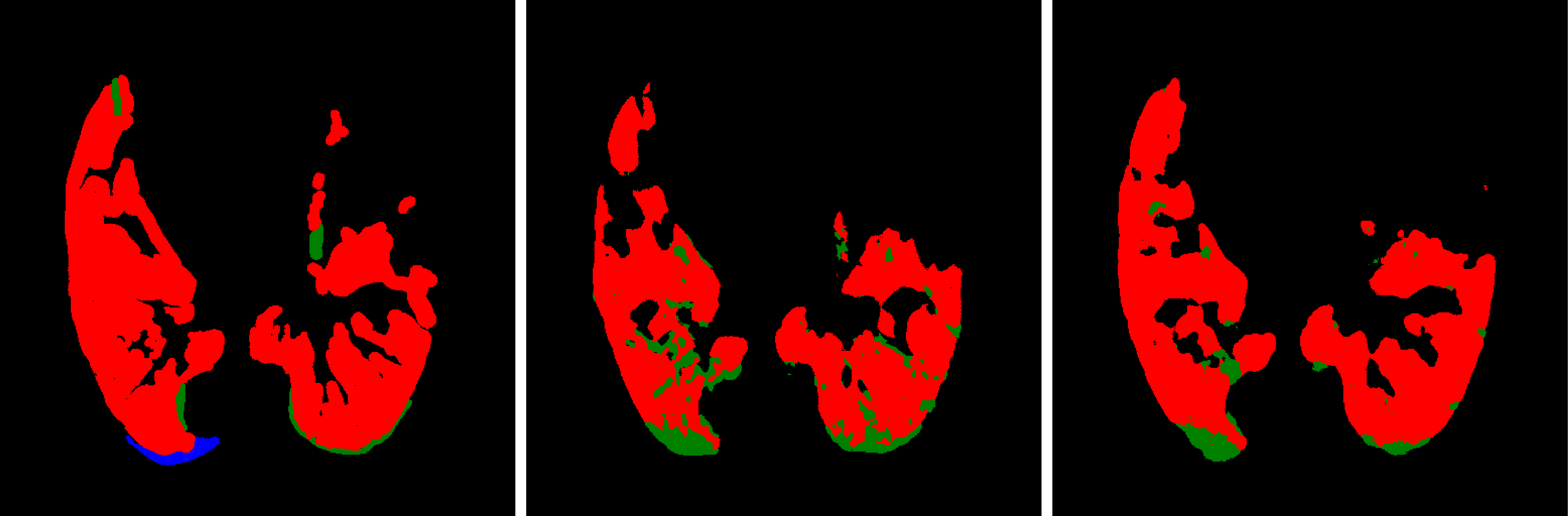}\\[.6ex]
\includegraphics[width=5.5in, height=1.7in]{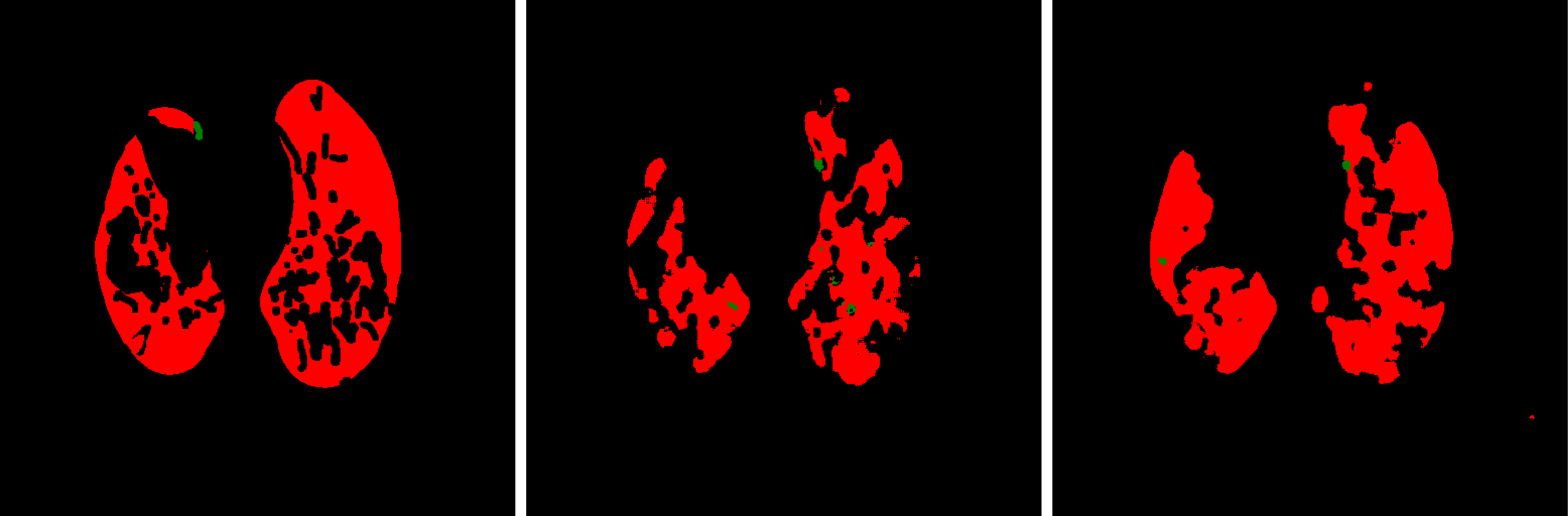}
\caption{Multi-class segmentation results on the COVID-19 CT Segmentation dataset: Ground truth (left); U-Net (center); and Sharp U-Net (right). Notice how Sharp U-Net is able to generate smoother predictions compared to U-Net.}
\label{Fig:multiclass}
\end{figure*}

We also test our proposed Sharp U-Net model in multi-class segmentation, which is more challenging than binary segmentation, using the COVID-19 CT Segmentation dataset. As expected, Sharp U-Net outperforms U-Net by 2.52\%. We visualize some multi-class segmentation results in Figure~\ref{Fig:multiclass}, which shows that Sharp U-Net is able to generate smoother predictions compared to U-Net.

\begin{table*}[!htb]
\caption{Evaluation results averaged over 5 folds for all datasets. We also report the standard deviation. Boldface numbers indicate the best segmentation performance. Notice that Sharp U-Net consistently outperforms U-Net.}
\label{Tab:results1}
\centering
\medskip
\begin{tabular}{@{}lcccccccccccc@{}}
\toprule
& \multicolumn{2}{c}{U-Net} & \multicolumn{2}{c}{Sharp U-Net} \\
\cmidrule(lr){2-3} \cmidrule(lr){4-5}
Dataset & Jaccard (\%) & Dice (\%) & Jaccard (\%) & Dice (\%) \\
\midrule
CVC-ClinicDB ~ & $70.02\pm 9.37$ &  $78.75\pm8.10$ & $\textbf{75.89}\pm2.06$ & $\textbf{83.65}\pm1.88$ \\
ISBI-2012 ~ & $83.95\pm3.47$ &  $91.23\pm2.11$ & $\textbf{87.00}\pm4.95$ & $\textbf{92.96}\pm2.95$ \\
COVID-19 CT Segmentation ~ & $85.24\pm1.96$ &  $91.85\pm 1.17$ & $\textbf{87.39}\pm3.07$ & $\textbf{93.01}\pm1.85$ \\
ISIC-2018 ~ & $74.74\pm4.46$ &  $83.25\pm3.73$ & $\textbf{76.07}\pm1.03$ & $\textbf{84.34}\pm0.86$ \\
Data Science Bowl 2018 ~ & $84.95\pm0.26$ &  $91.47\pm 0.18$ & $\textbf{85.26}\pm0.27$ & $\textbf{91.75}\pm0.15$ \\
Lung Segmentation ~ & $94.75\pm0.58$ &  $97.03\pm0.31$ & $\textbf{95.22}\pm0.67$ & $\textbf{97.25}\pm0.36$ \\
\bottomrule
\hline
\end{tabular}
\end{table*}

\begin{figure}[!htb]
\setlength{\tabcolsep}{.3em}
\centering
\begin{tabular}{cc}
\includegraphics[scale=.37]{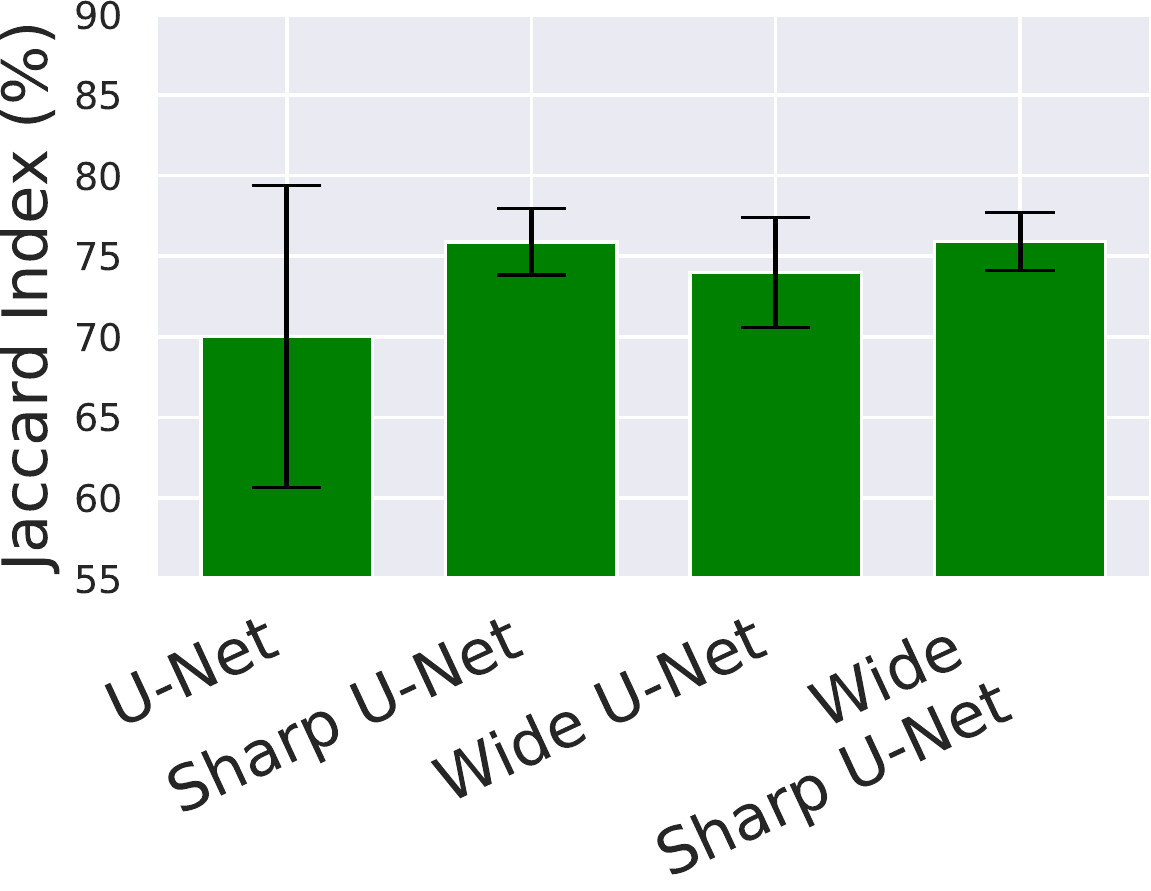} &
\includegraphics[scale=.37]{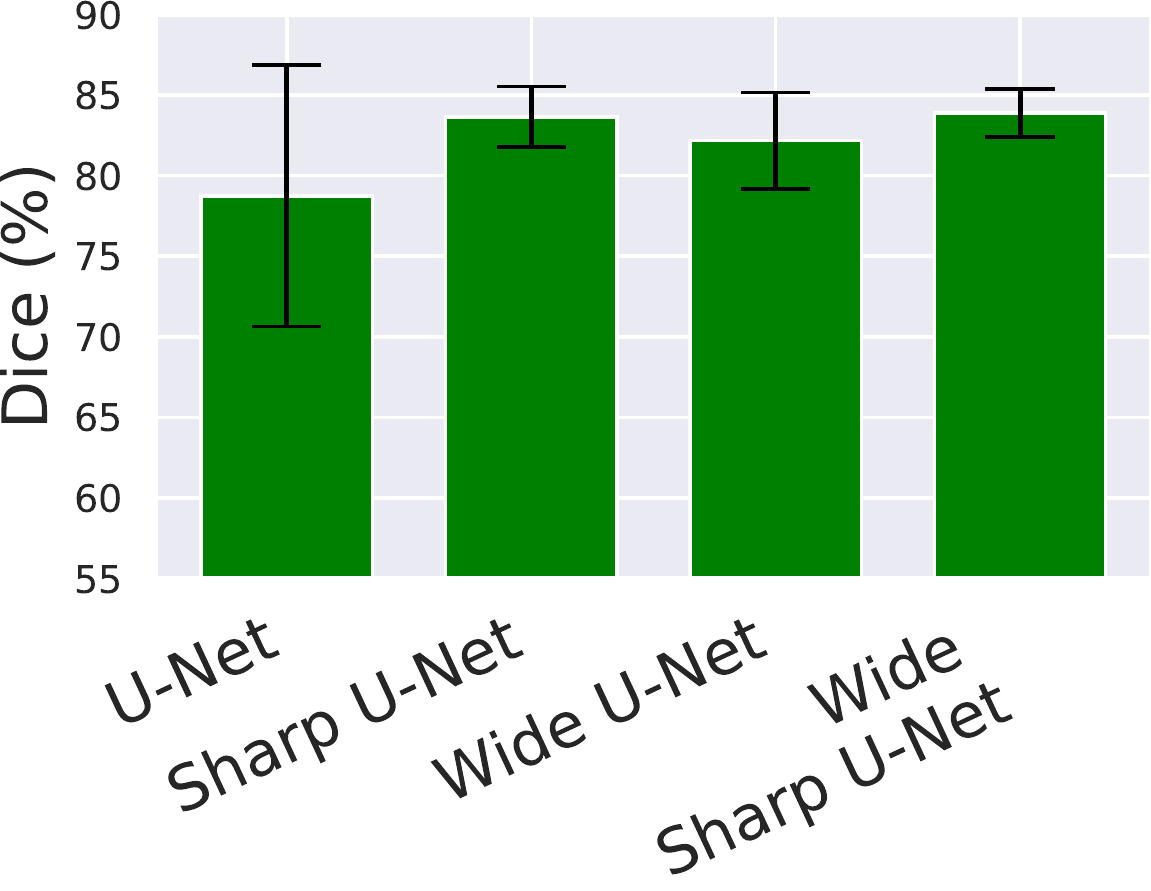}
\end{tabular}
\caption{Bar graphs with error bars in terms of Jaccard and Dice metrics (in percent) on the CVC-ClinicDB dataset using 5-fold cross-validation. For both metrics, Sharp U-Net performs better than Wide U-Net, albeit the latter has much more learnable parameters.}
\label{fig:barplots2}
\end{figure}

\medskip\noindent\textbf{Scaling Sharp U-Net improves performance.}\quad In Figure~\ref{fig:barplots2}, we illustrate the effectiveness of adding sharp blocks in comparison with adding traditional skip connections in Wide U-Net~\cite{zhou2019unet++} on the CVC-ClinicDB dataset for the task of polyp segmentation. Endoscopy images consist of homogeneous ROIs and background which makes it very challenging to differentiate between the two. Both U-Net and Sharp U-Net yield the lowest Jaccard and Dice scores among all the other datasets, showing the difficulty in segmenting polyps. From Figure~\ref{fig:barplots2}, it can be seen that when replacing the skip connections in Wide U-Net~\cite{zhou2019unet++} with sharp blocks, a relative improvement on the Jaccard index of 2.63\% is achieved. Also, the standard deviation of Wide Sharp U-Net is much lower than Wide U-Net (1.8024 and 3.4125, respectively). A similar pattern can be observed using the Dice coefficient. Another interesting finding from this experiment is that Sharp U-Net performs better than Wide U-Net, even though the latter has 9.1 million learnable parameters, whereas Sharp U-Net has only 7.8 million learnable parameters similar to U-Net.

Since recent works aim to improve performance on segmentation tasks by replacing the encoder subnetwork with pretrained convolutional networks such as VGG and ResNet, we also test the extensibility of training these pretrained models together with sharp blocks instead of the traditional skip connections. Table~\ref{Tab:results2} shows 5-fold cross-validation test results of U-Net and U-Nets with VGG~\cite{kalinin2020medical} and ResNet~\cite{he2016deep} encoders when training with skip connections and sharp blocks on the CVC-ClinicDB dataset. Notice that as the depth of the encoder increases, the Jaccard and Dice scores increase when trained with skip connection and sharp blocks. This is in line with findings reported in~\cite{kalinin2020medical}. As the encoder complexity increases, the performance gains when adding sharp blocks decreases. This is intuitive as the closer the predicted segmentation is to the perfect segmentation, the harder it is to improve it further, and can be largely attributed to the lower improvement of sharp U-Net as the complexity of encoder increases as compared to U-Net. When using the VGG encoder, it can be seen that training with sharp blocks yields almost comparable performance, but with sharp blocks better results are obtained. A significant reduction in the standard deviation can be observed (0.8090 vs. 3.6895) when using sharp blocks, suggesting that sharp blocks are more reliable compared to the traditional skip connections. The same pattern can be observed when using the ResNet~\cite{he2016deep} encoder, which consists of 32.7 million learnable parameters. When comparing Sharp U-Net with VGG Sharp U-Net, we can see that the performance is comparable, but the standard deviation of Sharp U-Net is much lower than that of VGG Sharp U-Net, demonstrating the effectiveness of sharp blocks. This is an interesting result as Sharp U-Net has three times fewer learnable parameters (7.8 million) compared to VGG Sharp U-Net (23.7 million).

\begin{table}[!htb]
\caption{Comparison results in terms of Jaccard and Dice metrics (in percent) on the CVC-ClinicDB dataset using 5-fold cross-validation tests.}
\label{Tab:results2}
	\centering
	\medskip
	\begin{tabular}{@{}lcccccccccccc@{}}
		\toprule
		& \multicolumn{2}{c}{CVC-ClinicDB} \\
		\cmidrule(lr){2-3}
		Model & Jaccard (\%) & Dice (\%) \\
		\midrule
		U-Net~\cite{ronneberger2015u}~ & $70.02 \pm 9.37$ & $78.75\pm8.10$ \\
		Sharp U-Net ~ & $75.89\pm2.06$ & $83.65\pm 1.88$ \\
		TernausNet-16 ~\cite{kalinin2020medical}~ & $76.07\pm3.69$ & $83.95\pm 2.95$ \\
		Sharp TernausNet-16 ~ & $76.33\pm0.81$ & $83.84\pm0.75$ \\
		U-Net + ResNet-50~\cite{diakogiannis2020resunet} ~ & $83.88\pm1.26$ & $89.83\pm1.25$ \\
		Sharp U-Net + ResNet-50 ~ & $\textbf{83.98}\pm0.27$ & $\textbf{90.05}\pm0.29$ \\
		\bottomrule
		\hline
	\end{tabular}
\end{table}

In Table~\ref{Tab:results3}, we report the performance comparison of Sharp U-Net and baseline models on all datasets. As can be seen, the use of sharp blocks significantly improves the segmentation performance of the proposed U-Net + ResNet-50 model, consistently outperforming baselines on all datasets for both binary and multi-class segmentation tasks.
\begin{table*}[!htb]
	\caption{Performance comparison of Sharp U-Net and baseline methods. Both evaluation metrics are averaged over 5 folds of cross-validation. We also report the standard deviation. Boldface numbers indicate the best segmentation performance.}
	\label{Tab:results3}
	\setlength{\tabcolsep}{3.8pt}
	\centering
	\medskip
	\resizebox{1\textwidth}{!}{%
	\begin{tabular}{@{}lcccccccccccc@{}}
		\toprule
		& \multicolumn{2}{c}{CVC-ClinicDB} & \multicolumn{2}{c}{ISBI-2012} & \multicolumn{2}{c}{COVID-19 CT Segmentation} & \multicolumn{2}{c}{ISIC-2018} & \multicolumn{2}{c}{Data Science Bowl 2018} & \multicolumn{2}{c}{Lung Segmentation}\\
		\cmidrule(lr){2-3} \cmidrule(lr){4-5} \cmidrule(lr){6-7} \cmidrule(lr){8-9} \cmidrule(lr){10-11} \cmidrule(lr){12-13}
		Model & Jaccard (\%) & Dice (\%) & Jaccard (\%) & Dice (\%) & Jaccard (\%) & Dice (\%) & Jaccard (\%) & Dice (\%) & Jaccard (\%) & Dice (\%) & Jaccard (\%) & Dice (\%) \\
		\midrule
		U-Net ~ & $70.02\pm  9.37$ &  $ 78.75 \pm 8.10$ & $83.95\pm3.47$ & $91.13\pm2.11$ & $85.24\pm  1.96$ &  $ 91.85 \pm 1.17$ & $74.74\pm  4.46$ &  $ 83.25 \pm 3.73 $ & $ 84.95\pm  0.26$ &  $ 91.47 \pm 0.18$ & $94.75\pm  0.58$ &  $ 97.03 \pm 0.31$ \\
        Wide U-Net ~ & $ 73.97 \pm 3.41$ &  $ 82.16 \pm 3.00$ & $ 85.34 \pm 2.99 $ & $ 91.54 \pm 2.05 $ & $ 86.22 \pm 1.80 $ & $92.33 \pm 1.10$ & $75.22 \pm 3.44$ & $83.88 \pm 3.10$ & $ 85.34 \pm 0.25$ & $91.50 \pm 0.17$ & $94.89 \pm 0.50$ & $97.40 \pm 0.30$\\
        TernausNet-16 ~ & $76.07 \pm 3.69$ &  $ 83.95 \pm 2.95$ & $ 86.31 \pm 2.22 $ & $ 92.58 \pm 1.77$ & $ 87.60 \pm 1.37 $ & $93.56 \pm 0.89$ & $76.32 \pm 2.37$ & $84.65 \pm 2.88$ & $ 86.30 \pm 0.25$ & $92.01 \pm 0.14$ & $95.41 \pm 0.47$ & $97.57 \pm 0.27$\\
        U-Net + ResNet-50 ~ & $83.88 \pm 1.26$ &  $ 89.83 \pm 1.25$ & $ 89.36 \pm 2.30 $ & $ 92.96 \pm 1.53$ & $ 89.79 \pm 1.55 $ & $94.04 \pm 0.81$ & $78.53 \pm 1.67$ & $85.78 \pm 1.54$ & $ 88.44 \pm 0.20$ & $93.10 \pm 0.13$ & $96.10 \pm 0.30$ & $98.12 \pm 0.21$\\
		\midrule
		Sharp U-Net (\textbf{Ours})~ & $75.89 \pm 2.06$ &  $ 83.65 \pm 1.88$ & $ 87.00 \pm 4.95 $ & $ 92.96 \pm 2.95$ & $ 87.39 \pm 3.07 $ & $93.01 \pm 1.85$ & $76.07 \pm 1.03$ & $84.34 \pm 0.86$ & $ 85.26 \pm 0.27$ & $91.75 \pm 0.15$ & $95.22 \pm 0.67$ & $97.25 \pm 0.36$\\
		Sharp U-Net + ResNet-50 (\textbf{Ours})~ & $\textbf{83.98} \pm 0.27$ &  $\textbf{90.05} \pm 0.29$ & $ \textbf{91.21} \pm 0.67 $ & $ \textbf{93.52} \pm 0.91$ & $ \textbf{91.22} \pm 1.20 $ & $ \textbf{94.65} \pm 0.69$ & $ \textbf{79.78} \pm 0.55$ & $ \textbf{87.01} \pm 0.90$ & $ \textbf{89.60} \pm 0.15$ & $\textbf{95.40} \pm 0.10$ & $\textbf{97.44} \pm 0.24$ & $\textbf{98.73} \pm 0.16$\\
		\bottomrule
		\hline
	\end{tabular}%
	}
\end{table*}

\begin{figure}[!htb]
\setlength{\tabcolsep}{.1em}
\centering
\begin{tabular}{cc}
\includegraphics[width=1.7in, height=1.2in]{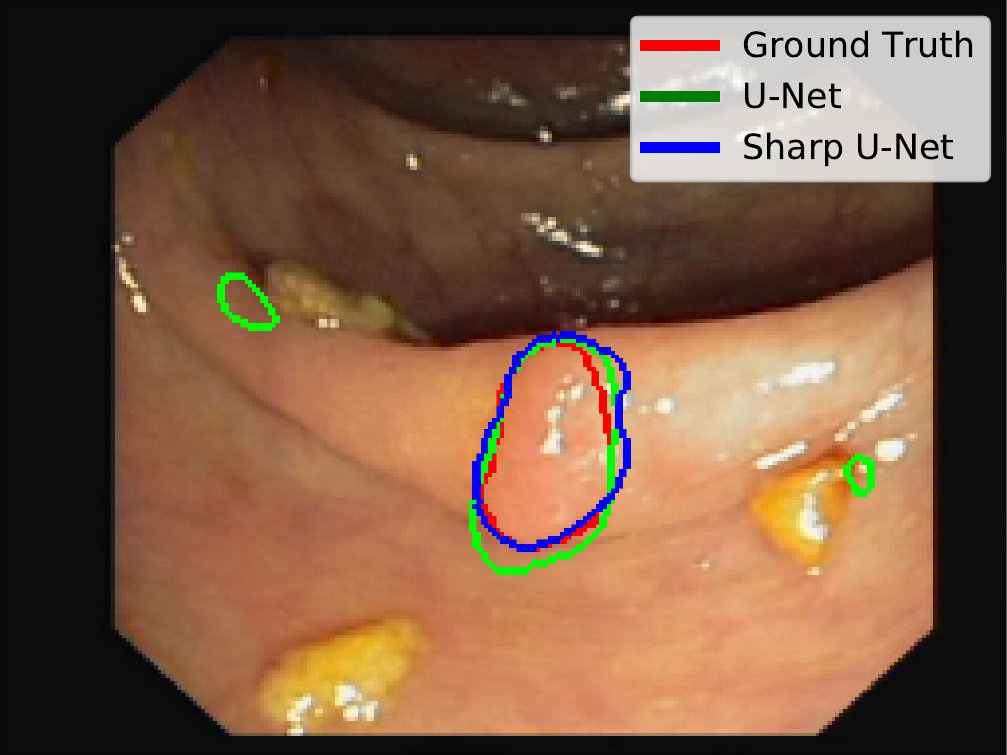} & \includegraphics[width=1.7in, height=1.2in]{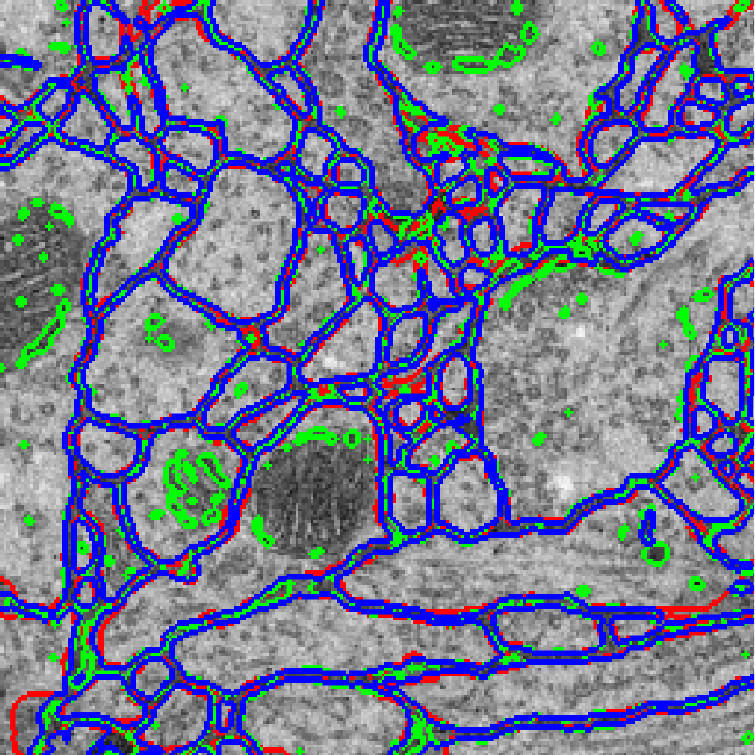} \\
(a) & (b) \\
\includegraphics[width=1.7in, height=1.2in]{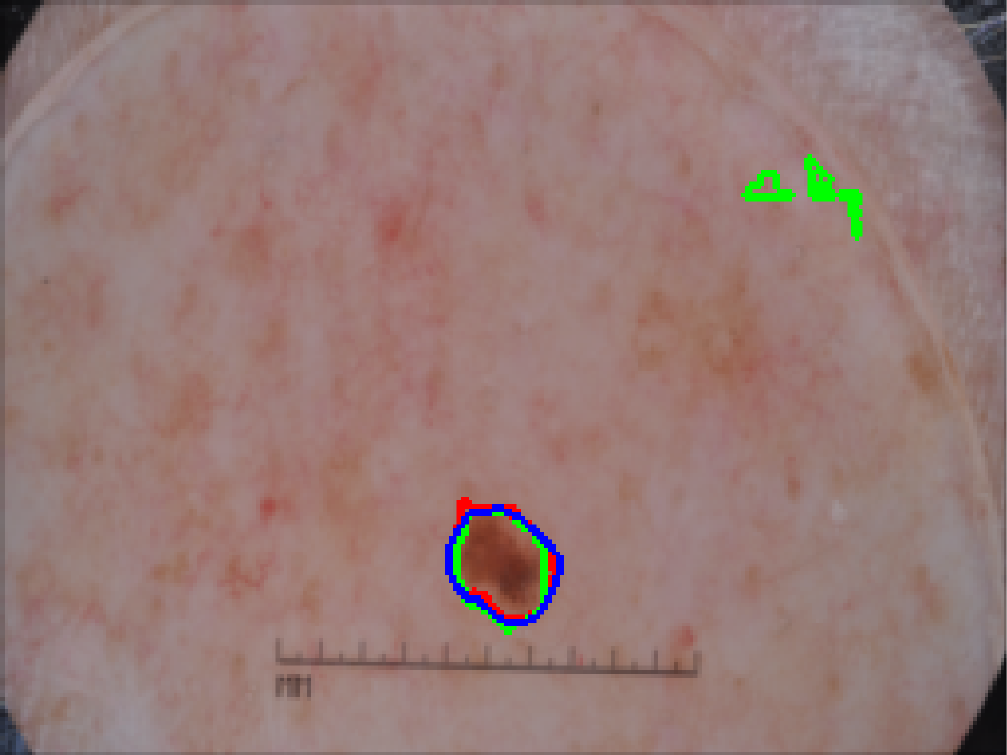} & \includegraphics[width=1.7in, height=1.2in]{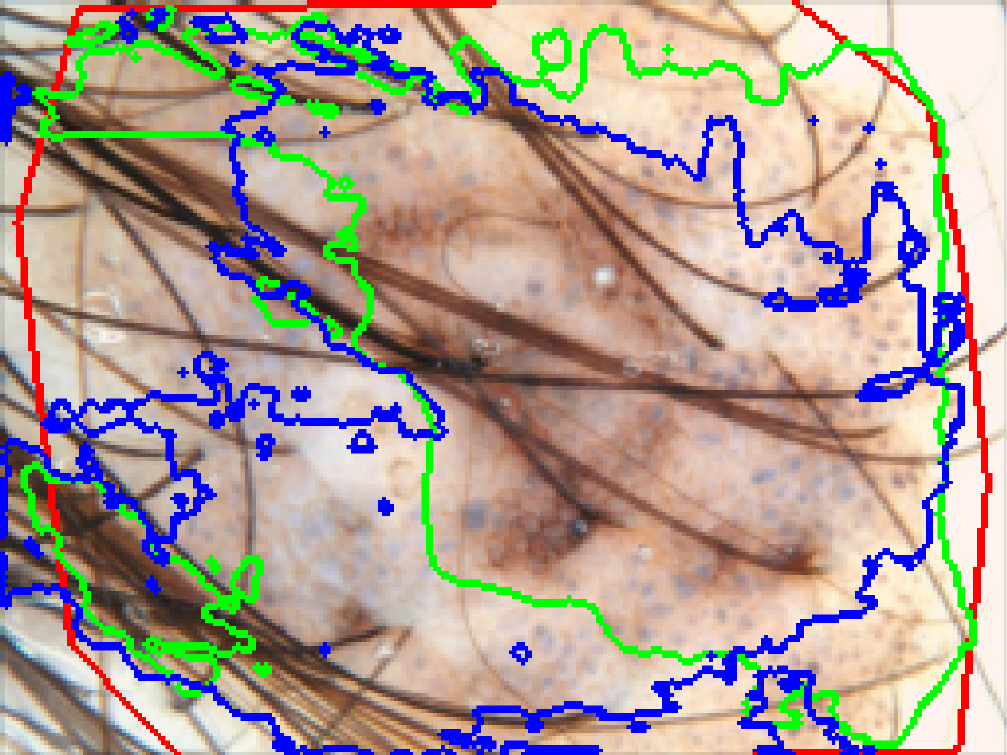} \\
(c) & (d) \\
\includegraphics[width=1.7in, height=1.2in]{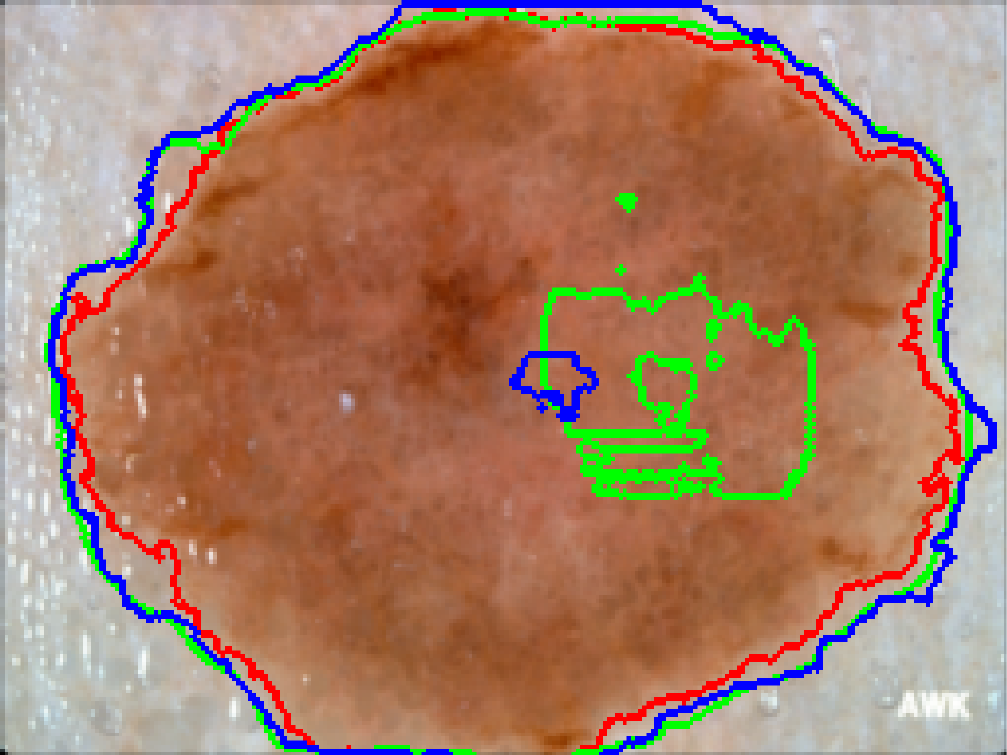} & \includegraphics[width=1.7in, height=1.2in]{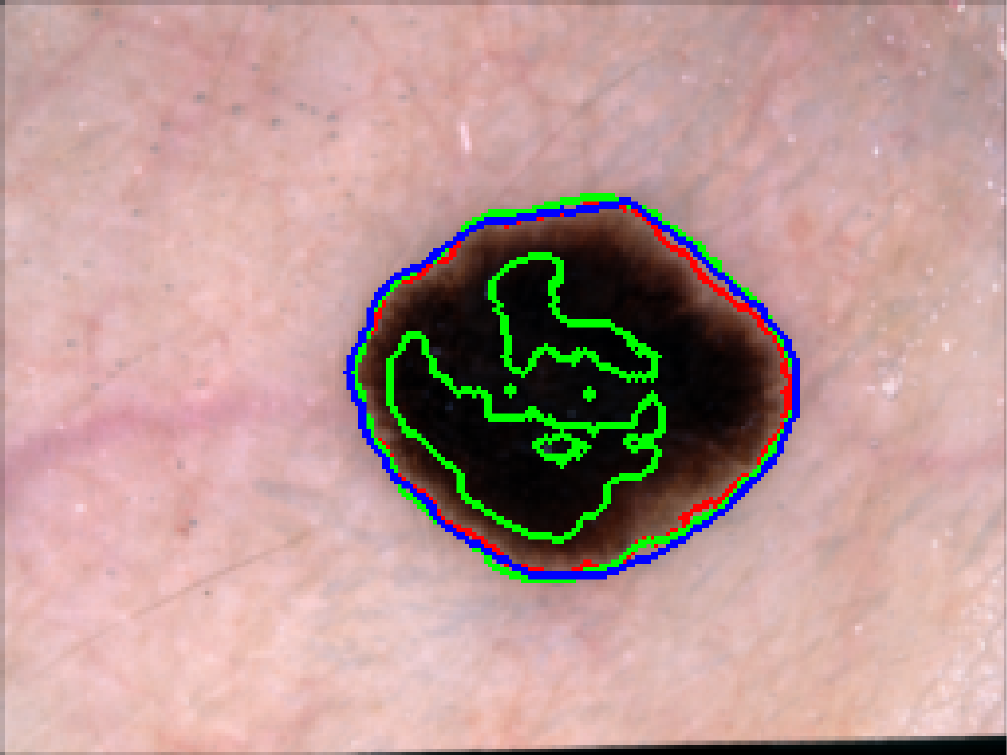} \\
(e) & (f)
\end{tabular}
\caption{Segmentation results of Sharp U-Net vs. U-Net on images with irregularities and hard-to-distinguish ROIs. Note that Sharp U-Net yields smoother predictions, while U-Net creates clusters of segmented regions in different parts of the images.}
\label{fig:1}
\end{figure}

\medskip\noindent\textbf{Sharp U-Net is robust to image irregularities.}\quad In Figure~\ref{fig:1}, we show qualitative results for some hard examples. These examples consist of segmenting ROIs that are occluded by artifacts such as hair or when the ROI and background are quite similar. While the segmented ROI predicted by Sharp U-Net is not perfect and kind of under-segmented, it yields substantially less amount of noise and fractured segmented ROIs compared to U-Net. For example, in  Figure~\ref{fig:1}(b), it can be observed that U-Net tends to segment non ROI regions in many areas of the image, whereas Sharp U-Net successfully segments most of the ROIs. Figures~\ref{fig:1}(e) and (f) show cases where U-Net tends to segment multiple non ROIs regions forming small clusters even though the ROI is easily distinguishable.

\begin{figure}[!htb]
\setlength{\tabcolsep}{.1em}
\centering
\begin{tabular}{cc}
\includegraphics[width=1.7in, height=1.2in]{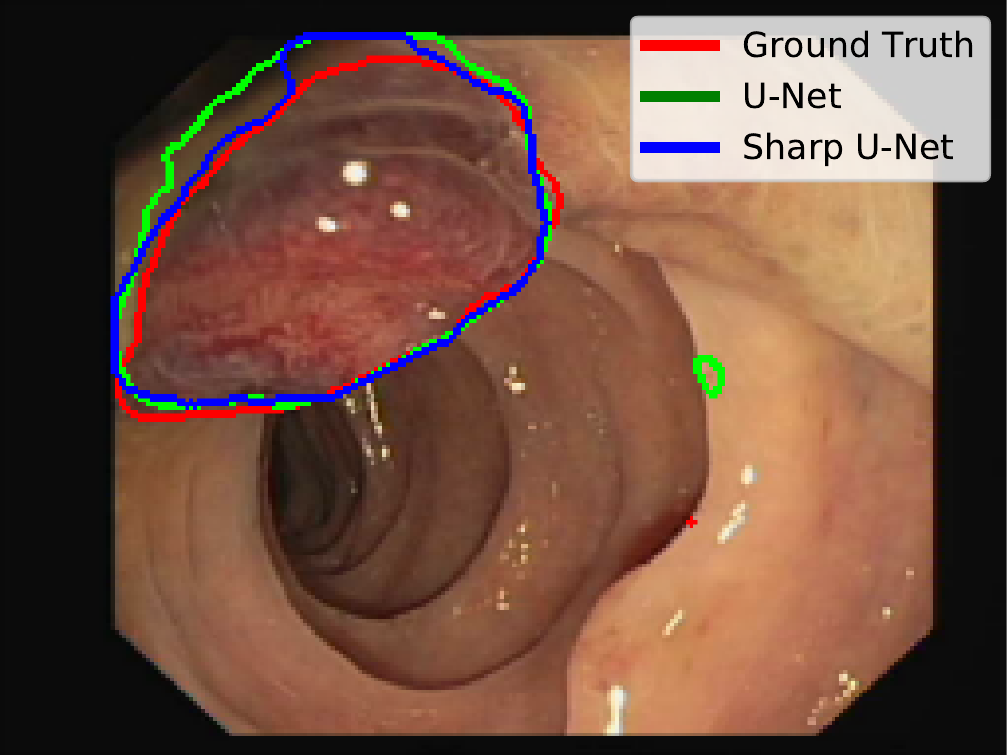} & \includegraphics[width=1.7in, height=1.2in]{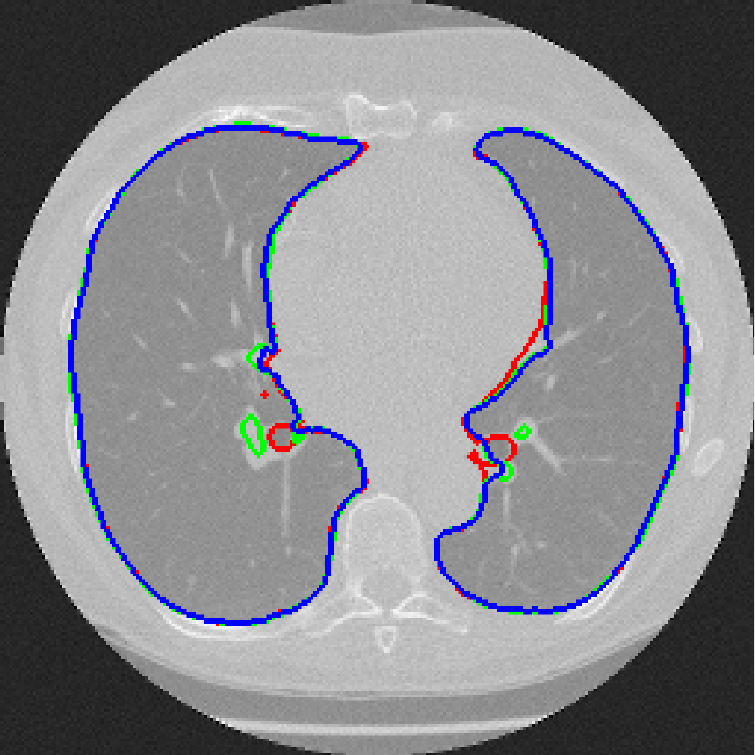} \\
(a) & (b) \\
\includegraphics[width=1.7in, height=1.2in]{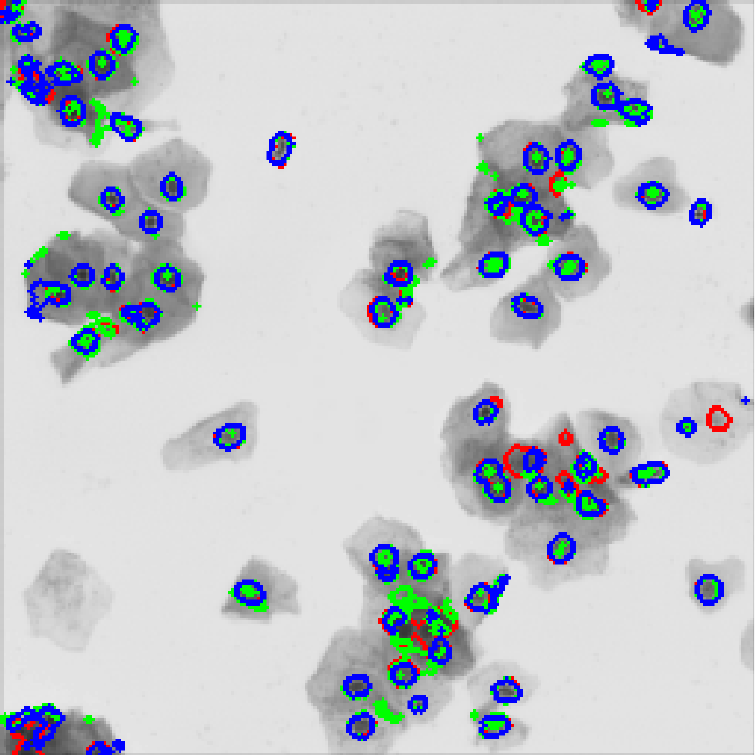} & \includegraphics[width=1.7in, height=1.2in]{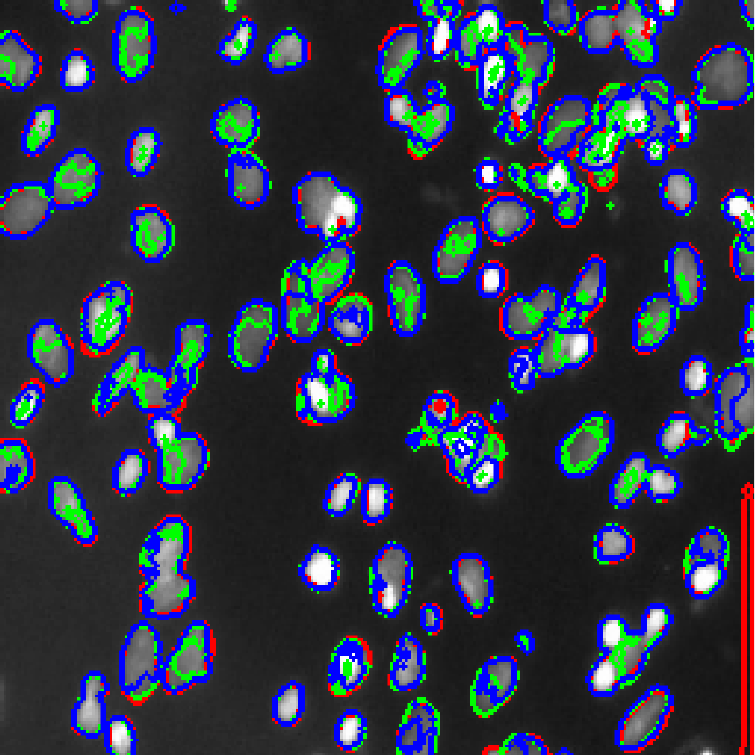} \\
(c) & (d) \\
\includegraphics[width=1.7in, height=1.2in]{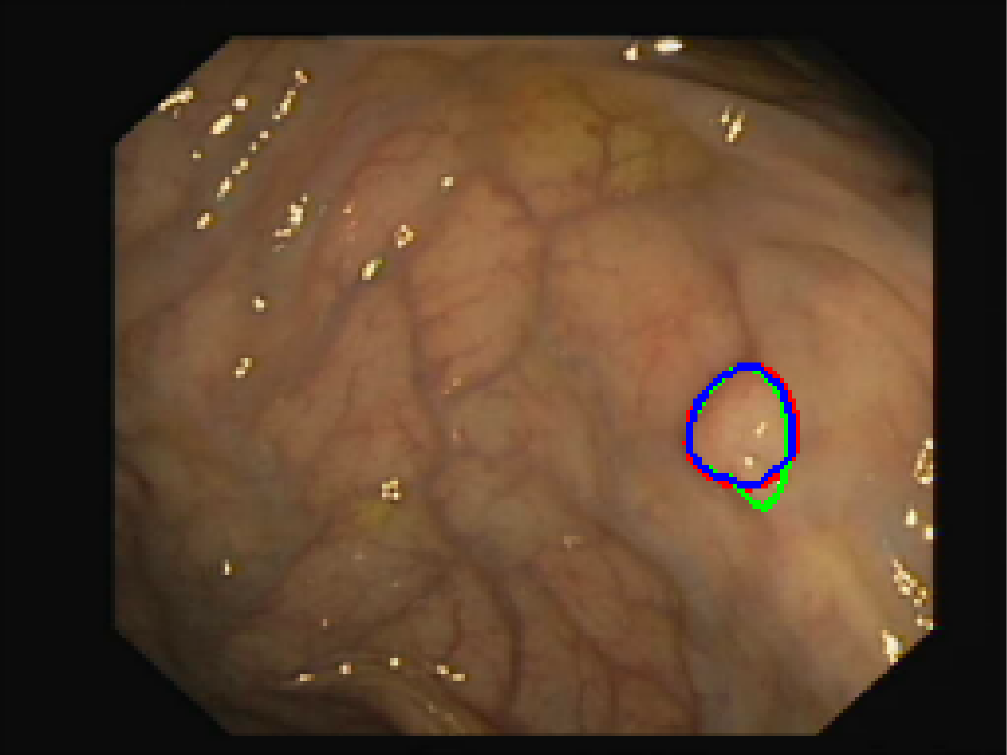} & \includegraphics[width=1.7in, height=1.2in]{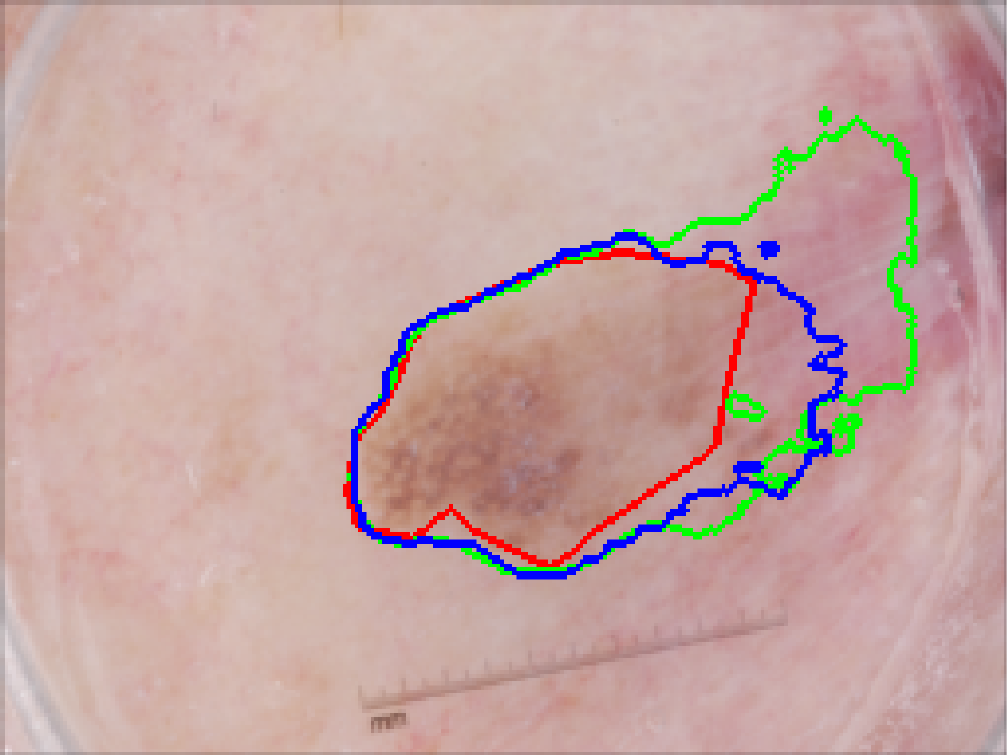} \\
(e) & (f)
\end{tabular}
\caption{Segmentation results of Sharp U-Net vs. U-Net on hard examples. Notice that U-Net suffers from over-segmentation even though in some cases the ROI is easily differentiable from the background, while Sharp U-Net yields a segmentation result that closely matches the ground truth.}
\label{fig:2}
\end{figure}

\begin{figure}[!htb]
\setlength{\tabcolsep}{.1em}
\centering
\begin{tabular}{cc}
\includegraphics[width=1.7in, height=1.2in]{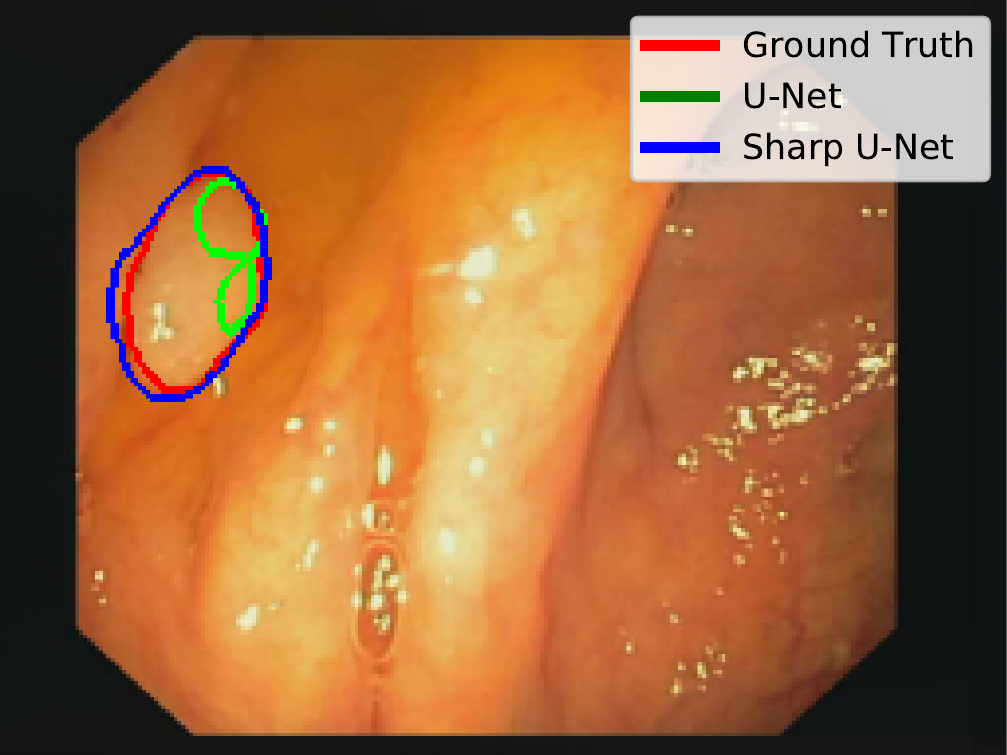} & \includegraphics[width=1.7in, height=1.2in]{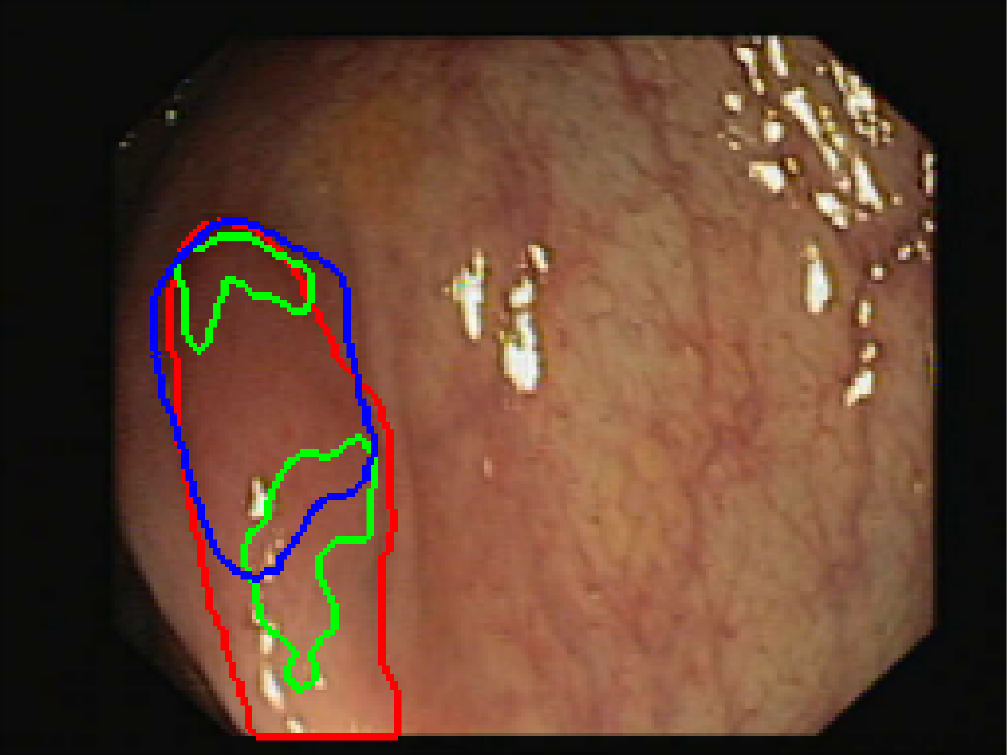} \\
(a) & (b) \\
\includegraphics[width=1.7in, height=1.2in]{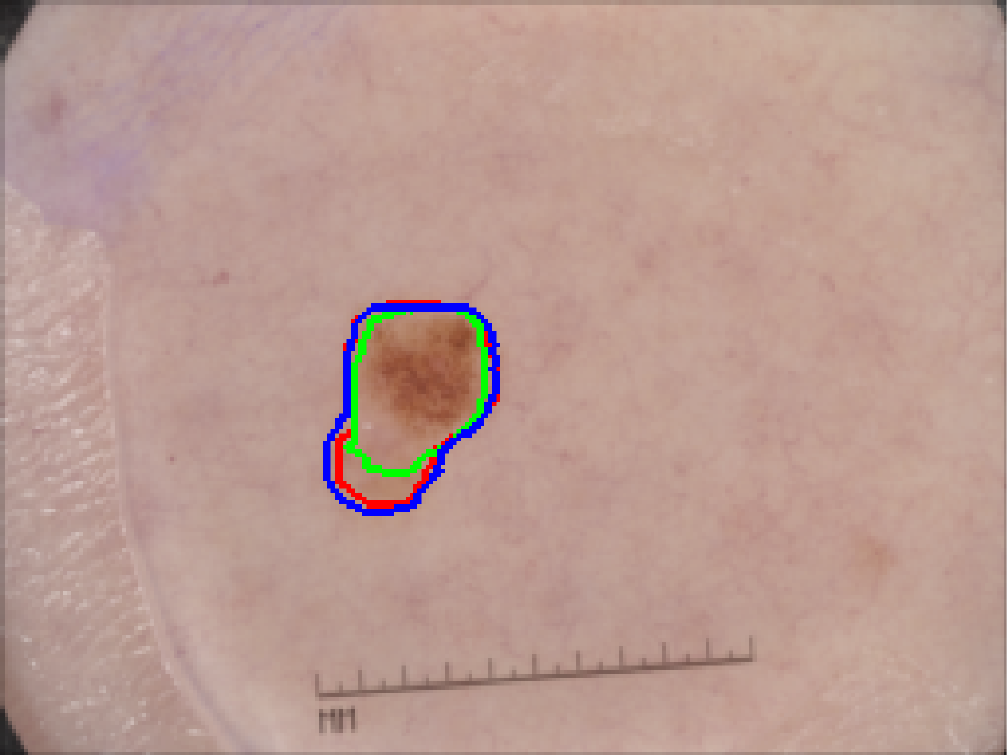} & \includegraphics[width=1.7in, height=1.2in]{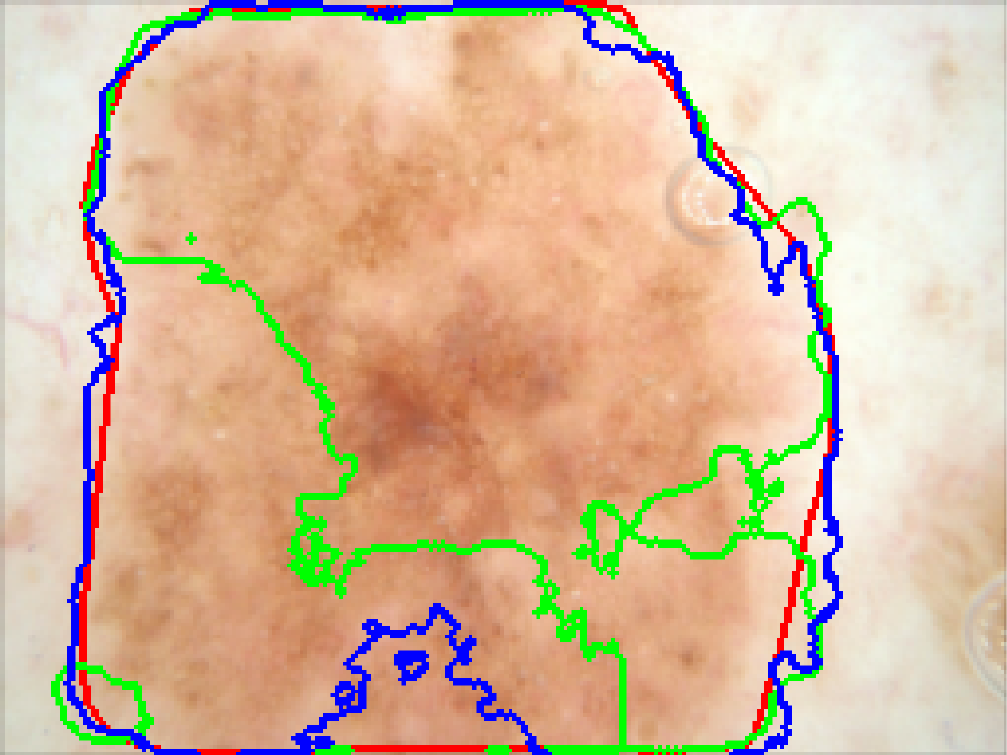} \\
(c) & (d) \\
\includegraphics[width=1.7in, height=1.2in]{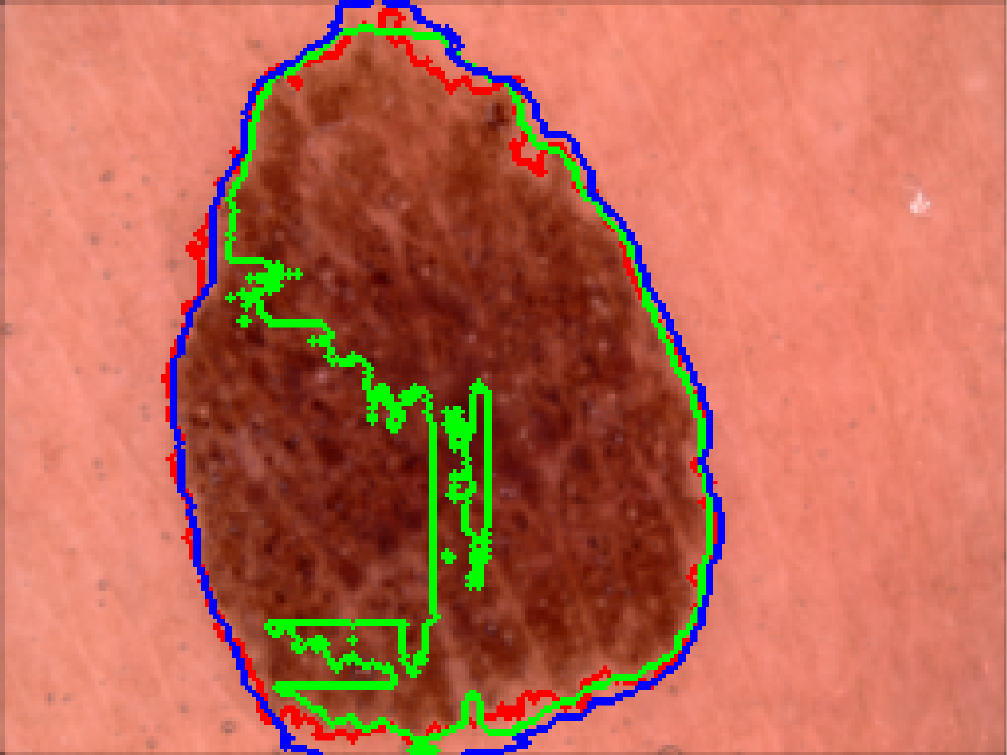} & \includegraphics[width=1.7in, height=1.2in]{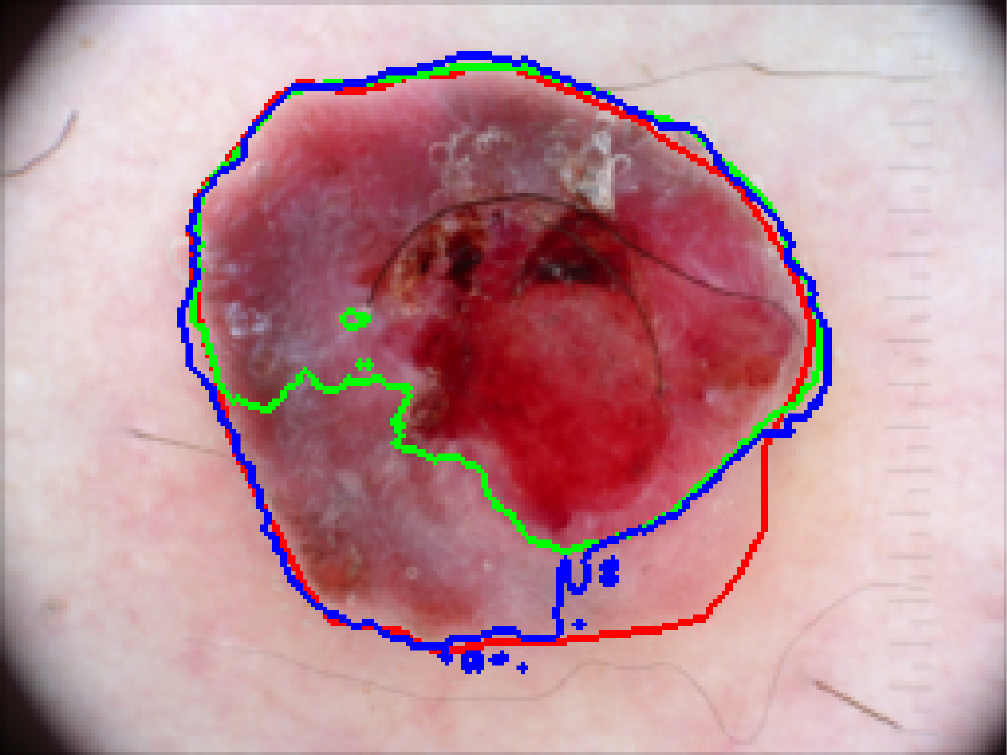} \\
(e) & (f)
\end{tabular}
\caption{Segmentation results in case where U-Net tends to under-segment the ROI. In some cases that are easily differentiable, U-Net performs better to some extent, but Sharp U-Net tends to predict the ROIs much better.}
\label{fig:3}
\end{figure}

\medskip\noindent\textbf{Mitigating under- and over-segmentation.}\quad In Figures~\ref{fig:2} and \ref{fig:3}, we illustrate some cases in which U-Net tends to over-segment or under-segment the ROIs. For example, we can see in Figure~\ref{fig:2}(b) that the U-Net tends to over-segment the ROIs even when there is a clear distinction between the foreground and background. More cases are shown in Figures~\ref{fig:2}(e) and (f), where the foreground is quite similar to the background. Even in such cases, Sharp U-Net is able to segment the ROIs better than U-Net, although Sharp U-Net tends to over-segment the ROIs in Figure~\ref{fig:2}(f). Figure~\ref{fig:3}(b) shows a low-contrast image from the CVC-ClinicDB dataset, and it can be seen that both Sharp U-Net and U-Net suffer from under-segmentation. However, Sharp U-Net is able to segment almost half of the ROIs, whereas U-Net forms two separate clusters. A similar pattern is observed in Figures~\ref{fig:3}(d) and (f), which show cases from the ISIC-2018 dataset for the task of segmenting lesions. Figures~\ref{fig:3}(a), (c) and (e) show cases, where Sharp U-Net yields near perfectly segmentation of the ROIs, even with almost no difference between the foreground and background image in Figure~\ref{fig:3}(a). This suggests that Sharp U-Net is able to achieve better performance, while tackling the over- and under-segmentation problems.

\subsection{Feature Visualization}
Understanding and interpreting the predictions made by a deep learning model provides valuable insights into the input data and the features learned by the model so that the results can be easily understood by human experts. In order to visually explain the decisions made by the proposed segmentation architecture and baseline methods for segmenting the ROIs, we use the gradient-weighted class activation map (Grad-CAM)~\cite{selvaraju2017grad} to generate the saliency maps, which highlight the most influential features affecting the predictions. Since convolutional features retain spatial information and that each feature value indicates whether the corresponding visual pattern exists in its receptive field, the output of the convolutional layer of the network shows the discriminative regions of the image.

\begin{figure}[!htb]
\centering
\includegraphics[width=3.5in]{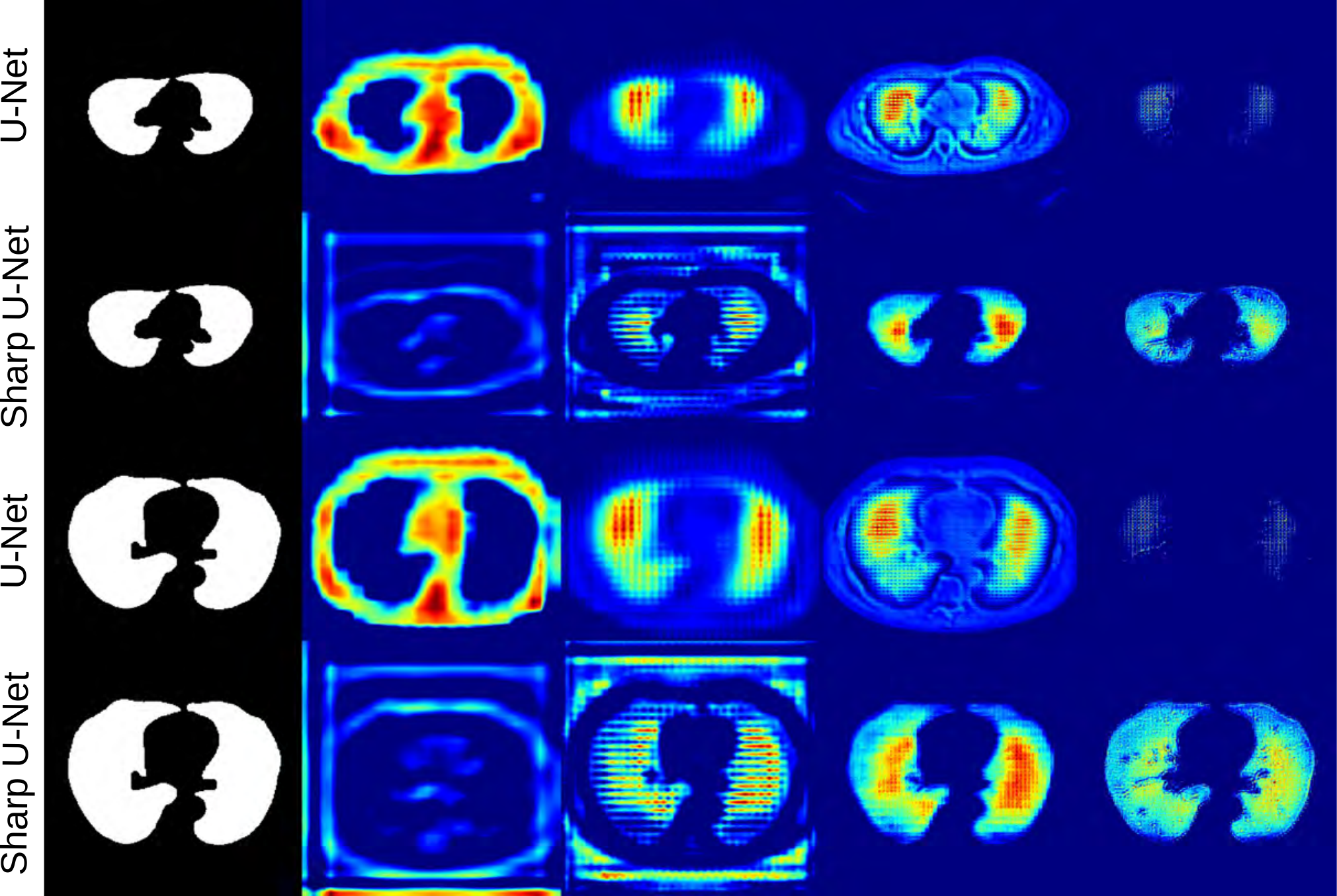}
\caption{GradCAM heatmaps showing high activations for the regions affecting the predicted segmentation. For comparison between skip connections and sharp blocks, we visualize the upsampling layers of the decoder subnetwork. From left to right; ground truth and first to fourth concatenation layers.}
\label{Fig:gradcam}
\end{figure}

In Figure~\ref{Fig:gradcam}, we visualize the class activation maps of the upsampling layers of two lung segmentation cases for both Sharp U-Net and U-Net. As can be seen, U-Net demonstrates high activations around the ROIs on the upsampling layer (i.e. first concatenation layer), which is the shallowest skip connection bridging the encoder decoder subnetworks. As the skip connections get deeper (i.e. second and third concatenation layers), the network shows high activations on the ROIs. For the deepest upsampling layer (i.e. fourth concatenation layer), lightly noticeable activations for the ROIs are observed. As shown in Figure~\ref{Fig:gradcam}, the opposite scenario is observed when using Sharp U-Net. It can be seen that the early upsampling layers show barely perceptible activations for the ROIs compared to U-Net, but exhibit high activations for the deeper upsampling layer, which shows high activations similar to the ground truth mask. Ideally, after the last upsampling layer operation and since the network outputs the final predictions, we should expect high activations for the ROIs similar to the ground truth mask. This better performance of Sharp U-Net suggests that the sharp blocks are able to better fuse the encoder-decoder features, yielding the merging of less semantically dissimilar features. This, in turn, enables the network to learn better representations of the input data and segment ROIs better than U-Net.

\subsection{Discussion}
We assess the effectiveness of Sharp U-Net on both binary and multi-class segmentation of medical images from multiple modalities, namely electron microscopy (EM), endoscopy, dermoscopy, nuclei and computed tomography (CT). The proposed Sharp U-Net outperforms the U-Net model on all tasks with improvements of 12.6\%, 3.63\%, 2.52\%, 1.79\%, 0.48\% and 0.63\% on the CVC-ClinicDB, ISBI-2012, COVID-19 CT Segmentation, ISIC-2018, Data Science Bowl 2018, and Lung Segmentation, respectively. An improvement of 12.6\% is achieved on the CVC-ClinicDB dataset of endoscopy images. This is significant because in addition to diverse shape, size and structure, the foreground and background are very similar; thereby it becomes a challenging task to differentiate between images. The second best improvement of 3.63\% is achieved on the ISIB-2012 dataset. Interestingly, this dataset consists of pixel-wise labels of the region of interests that are the majority of the images. This is very rare because in most labeled medical image segmentation datasets, the pixel-wise labels occupy only a very small fraction of the entire image. This poses a different challenge due to the majority of labels being foreground, where models would tend to over-segment the images in order to minimize the loss during optimization. This issue is quite similar to the imbalanced image classification problem, where a model tends to classify the over-represented class samples. Even in this rare case, we find that Sharp U-Net can provide more visually accurate results. Therefore, we believe that our proposed architecture can serve as a viable segmentation tool for binary and multi-class segmentation of medical images.

\section{Conclusion}
In this paper, we introduced Sharp U-Net, a new encoder-decoder depthwise fully convolutional network architecture for binary and multi-class segmentation. The core idea is to convolve the output of the encoder feature map with a sharpening spatial filter prior to performing fusion with the decoder features. The proposed segmentation framework is not only able to make the encoder and decoder features  semantically less dissimilar, but also helps smooth out artifacts throughout the network layers during the early stages of training due to untrained parameters. Experimental results demonstrate that our proposed architecture consistently outperforms or matches the state-of-the-art baselines on various benchmarks for binary and multi-class segmentation on biomedical images from different modalities. More importantly, our approach achieved significant performance improvements without adding any extra learnable parameters. In addition, we showed that Sharp U-Net can be easily scaled for improved performance, outperforming baselines that have three times the number of learnable parameters. For future work, we plan to explore effective techniques for handling the semantic gap between the encoder and decoder features, we well as to extend our approach to volumetric medical image segmentation.

\bibliographystyle{ieeetr}
\bibliography{references}
\end{document}